\documentclass[11pt]{article}
\usepackage{amssymb}
\usepackage{amsmath}
\usepackage{graphicx}
\usepackage{color}

\renewcommand{\vec}[1]{{\bf #1}}       
\def\beq{\begin{eqnarray}}    
\def\eeq{\end{eqnarray}}      

\newcommand{\OMo}{\Omega_{M}^0}

\newcommand{\OLo}{\Omega_{\Lambda}^0}

\newcommand{\rc}{\rho_c}
\newcommand{\rco}{\rho_{c}^0}
\newcommand{\rmo}{\rho_{m}^0}

\newcommand{\rM}{\rho_M}
\newcommand{\rmr}{\rho_m}
\newcommand{\pmr}{p_m}
\newcommand{\rMo}{\rho_{M}^0}

\newcommand{\rR}{\rho_R}
\newcommand{\rD}{\rho_D}

\newcommand{\rX}{\rho_X}
\newcommand{\rB}{\rho_B}
\newcommand{\rLep}{\rho_L}

\newcommand{\wm}{\omega_m}

\newcommand{\rL}{\rho_{\CC}}

\newcommand{\rLo}{\rho_{\CC}^0}
\newcommand{\pD}{p_D}
\newcommand{\wD}{\omega_D}

\newcommand{\wL}{\omega_{\CC}}
\newcommand{\CC}{\Lambda}

\newcommand{\LQCD}{\Lambda_{\rm QCD}}
\newcommand{\HB}{\cal H}
\newcommand{\mupe}{\mu_{\rm pe}}
\newcommand{\OMB}{\Omega_B}
\newcommand{\OMBo}{\Omega^0_B}
\newcommand{\ODM}{\Omega_{\rm DM}}
\newcommand{\ODMo}{\Omega^0_{\rm DM}}
\newcommand{\nuQCD}{\nu_{\rm QCD}}
\newcommand{\nuX}{\nu_{X}}
\newcommand{\nuB}{\nu_{B}}
\newcommand{\nueff}{\nu_{\rm eff}}

\newcommand{\mysection}[1]{\section{#1}
\renewcommand{\theequation}{\thesection.\arabic{equation}}
\setcounter{equation}{0}}

\newcommand{\mysubsection}[1]{\subsection{#1}
\renewcommand{\theequation}{\thesubsection.\arabic{equation}}
\setcounter{equation}{0}}

\begin{document}



 \hyphenation{cos-mo-lo-gi-cal
sig-ni-fi-cant par-ti-cu-lar}




\begin{center}
{\LARGE \textbf{Matter Non-conservation in the Universe\\ and Dynamical
Dark Energy}} \vskip 2mm

 \vskip 8mm

\textbf{Harald Fritzsch\,$^{a}$,  Joan Sol\`{a}\,$^{b}$}

\vskip0.5cm

$^{a}$ Physik-Department, Universit\"at M\"unchen, D-80333 Munich,
Germany, and

Institute for Advanced Study, Nanyang Technological University, Singapore.

\vspace{0.5cm}

$^{b}$ High Energy Physics Group, Dept. ECM and Institut de Ci{\`e}ncies del Cosmos\\
Univ. de Barcelona, Av. Diagonal 647, E-08028 Barcelona, Catalonia, Spain

\vspace{0.25cm}

E-mails: fritzsch@mppmu.mpg.de, sola@ecm.ub.es \vskip2mm

\end{center}
\vskip 15mm

\begin{quotation}
\noindent {\large\it \underline{Abstract}}.\ \ In an expanding universe
the vacuum energy density $\rL$ is expected to be a dynamical quantity. In
quantum field theory in curved space-time $\rL$ should exhibit a slow
evolution, determined by the expansion rate of the universe $H$. Recent
measurements on the time variation of the fine structure constant and of
the proton-electron mass ratio suggest that basic quantities of the
Standard Model, such as the QCD scale parameter $\LQCD$, may not be
conserved in the course of the cosmological evolution. The masses of the
nucleons $m_N$ and of the atomic nuclei would also be affected. Matter is
not conserved in such a universe. These measurements can be interpreted as
a leakage of matter into vacuum or vice versa. We point out that the
amount of leakage necessary to explain the measured value of
$\dot{m}_N/m_N$ could be of the same order of magnitude as the
observationally allowed value of $\dot\rho_{\CC}/\rL$, with a possible
contribution from the dark matter particles. The dark energy in our
universe could be the dynamical vacuum energy in interaction with ordinary
baryonic matter as well as with dark matter.
\end{quotation}
\vskip 8mm

PACS numbers:\ {95.36.+x, 04.62.+v, 11.10.Hi}

\newpage

\vskip 6mm

 \noindent \mysection{Introduction}
 \label{Introduction}



The Standard Model (SM) of the strong and electroweak (EW) interactions
contains 27 independent fundamental constants: the QED fine structure
constant $\alpha_{\rm em}=e^2/4\pi$, the $SU(2)_L$ gauge coupling $g$ of
the EW interactions, the gauge coupling constant of the strong
interactions $g_s$, the mass $M_W$ of the weak gauge boson $W$, the mass
$M_{\HB}$ of the Higgs boson $\HB$, the 12 masses of the quarks and
leptons, the 3 mixing angles of the quark mass matrix, a CP-violating
phase, the $3$ mixing angles in the lepton sector, a CP-violating phase
and two additional phases, if the neutrino masses are Majorana masses. One
of the parameters in the list, the mass of the Higgs boson $M_{\HB}$, has
not been measured thus far, despite some recent
hints\,\cite{HiggsFinding}.

If we include the Einstein-Hilbert (EH)
Lagrangian of gravity, there are two more fundamental constants, both of them
dimensionful: Newton's gravitational coupling $G_N$ and the cosmological
constant $\CC$ (also denoted as the CC term). The gravity constant has the
dimension of an inverse mass squared (in natural units): $G=1/M_P^2$,
where $M_P\simeq 1.22\times 10^{19}$ GeV is the Planck mass, the largest
mass scale in the universe. The cosmological constant has the dimension
of mass squared, the mass being of order $H_0\sim 10^{-33}$ eV, i.e.
essentially the value of the Hubble parameter at present (the smallest
mass scale in the universe).

Until recently the observational data on $|\CC|$ could only place an
upper bound, but now cosmological observations give a value, which
is tiny, but non-vanishing (in particle
physics standards) and positive\,\cite{WMAP,SNIa}. It can be
expressed as an energy density: $\rLo=\CC/(8\pi\,G_N)\sim 10^{-47}$
GeV$^4$ -- the so-called vacuum energy density. We can define
the mass scale associated to the CC term as follows:
$m_{\CC}\equiv\left(\rLo\right)^{1/4}\sim 10^{-3}$ eV. The scale
$m_{\CC}$ is the geometric mean of the two
extreme mass scales in the universe: $m_{\CC}\sim
\left(H_0\,M_P\right)^{1/2}$. In the $\CC$CDM model (i.e.
the standard model of cosmology) this scale associated to the vacuum
is assumed to be constant. This is a big puzzle within the $\CC$CDM model.

The dark energy (DE) problem was originally presented as the
cosmological constant (CC)
problem\,\cite{WeinbergRMP,PeeblesRatra03,CCRev}. This is one the
basic problems of physics, ever since it was first formulated
45 years ago\,\cite{Zeldovich67}\,\footnote{For a recent detailed account
of the old fine tuning CC problem, see e.g. sect.~2 and Appendix~B
of\,\cite{RelaxTH}.}.

In this paper we suggest the possibility that some of the cosmological
constant problems might be related to basic parameters
of the Standard Model. The nucleon mass and
the QCD scale $\LQCD$ might not have remained constant throughout the
history of the cosmological
evolution\,\cite{FritzschBarcelona11,Fritzsch11,Fritzsch06,Langacker02}.
This is related to the time variation of the fine structure constant.
Constraints on the ratio $\dot{\alpha}_{\rm
em}/\alpha_{\rm em}$ can be derived from limits on the position of
nuclear resonances in natural fission reactors which have been
working for the last few billions years -- the so-called ``Oklo
phenomenon''\,\cite{Shlyakhter1976,DamourDyson96,Fujii2000}.

There could also be a cosmic time variation of the strong coupling
constant, $\alpha_s$, related to the variation of the fundamental QCD
scale $\LQCD$. One expects that $\dot{\Lambda}_{\rm QCD}/\LQCD$ should be
larger than $\dot{\alpha}_{\rm em}/\alpha_{\rm em}$. Recent high precision
experiments performed both in the laboratory with atomic
clocks\,\cite{Haensch12,Haensch04,AtomicClocks} and in astrophysics using
data from quasars\,\cite{Reinhold06} support these ideas.

It has been suggested that the parameters of the EH action, $G_N$ and
$\CC$, may be varying with time due to the interaction of the vacuum
with the matter\,\footnote{For a recent review, see
e.g. \cite{JSP11a} and references therein.}. This time evolution might be
linked to the time variation of the QCD scale and to the time shift of all
the particle masses, including the dark matter ones. All the atomic masses
of the chemical elements would be affected.

Here is the outline of this paper. In section \ref{sect:timeCC} we review
the models with time evolving cosmological parameters. In section
\ref{sect:runningCC} we specialize to a class of these models, where the
time evolution is viewed as a renormalization group evolution. In section
\ref{sect:timeMasses} we describe some experiments, providing evidence of
the time variation of masses and couplings, and suggests a link of this
variation with that of the cosmological parameters. In section
\ref{sect:leakage} we propose the existence of a leakage of matter into
the vacuum as a possible source of dynamical dark energy and compute the
variation of the particle masses and the QCD scale with the Hubble rate.
The last section contains our conclusions.

\section{Cosmological models with time evolving parameters} \label{sect:timeCC}

We discuss the possibility that the cosmic time variations of the constants
of particle physics and of cosmology are related. Scenarios, in which $G$
could be variable, have been previously discussed in the literature. Dirac
suggested in the thirties (through his ``large number
hypothesis''\,\cite{Dirac37}), that the gravitational constant $G$ could be
varying with time in correlation with other fundamental constants. We also
mention the ideas on time varying fundamental constants by Milne and
Jordan at about the same time\,\cite{MilneJordan}. Later the time
variation of $G$ was tied to the
existence of a dynamical scalar field coupled to the curvature - the
original Jordan and Brans-Dicke proposals\,\cite{JBD}.

Consider the General Relativity field equations in the presence of the
cosmological term:
\begin{equation}
G_{\mu\nu}-\,g_{\mu\nu}\CC=8\pi G\,T_{\mu\nu}\,. \label{EE}
\end{equation}
Here $G_{\mu\nu}=R_{\mu \nu }-\frac{1}{2}g_{\mu \nu }R$ is the Einstein
tensor, and $T_{\mu\nu}$ is the energy-momentum tensor of the isotropic
matter and radiation in the universe. Without violating the
Cosmological Principle within the context of the FLRW (Friedmann-Lema\^\i
tre-Robertson-Walker) cosmology, nothing prevents the parameters
$G=G(t)$ and $\CC=\CC(t)$ to be functions of the
cosmic time, as it is the case with the scale factor itself $a=a(t)$. The
possibility of a variable CC term has been considered by many authors from different
points of view\,\cite{CCvariable,OldScalar}, including the more recent
quintessence approach -- cf.\,\cite{PeeblesRatra03} and references
therein.

The contribution from the $\CC$ term, originally on the \textit{l.h.s.} of
Einstein's equations, can be absorbed on the \textit{r.h.s.} after
introducing the quantity $\rL=\CC/(8\pi G_N)$, which represents the vacuum
energy density associated to the cosmological term. Einstein's equations
can then be rewritten formally the same way as in (\ref{EE}), but replacing
the ordinary energy-momentum tensor of matter by the total energy-momentum
tensor of matter and the vacuum energy:
\begin{equation} \label{tildeEMT}
{T}_{\mu\nu}\to \tilde{T}_{\mu\nu}\equiv T_{\mu\nu}+g_{\mu\nu}\,\rL
=  (\rL- p_m)\,g_{\mu\nu}+\big(\rmr+\pmr\,\big)\,U_{\mu}\,U_{\nu}\,.
\end{equation}
Here  $\rmr$ and $\pmr$ are the proper density and pressure of the
isotropic matter, and $U_{\mu}$ is the $4$-velocity of the cosmic fluid.

The corresponding equation of state
(EoS) $\wm=\pmr/\rmr$ reads: $\wm=1/3$ and $\wm=0$, for relativistic and
non-relativistic matter respectively. The redefinition of the
energy-momentum tensor can be done in the same way as in
Eq.\,(\ref{tildeEMT}), whether $\rL$ is strictly constant or time varying.
In both cases it enters with the equation of state $p_{\CC}=-\rL$, i.e.
$\wL=-1$. This is in distinction to the general DE fluids, whose EoS
take the generic form $\pD=\wD\rD$ (with
$\wD<-1/3$)\,\cite{PeeblesRatra03,CCRev}.

We discuss now some possible scenarios for variable cosmological
parameters that appear when we solve Einstein's equations (\ref{EE}) in
the spatially flat FLRW metric, $ds^{2}=dt^{2}-a^{2}(t)d\vec{x}^{2}$,
where $a(t)$ is the time-evolving scale factor.  We restrict ourselves to
the spatially flat case, since this seems to be the most plausible
possibility in view of the present observational data\,\cite{WMAP} and the
natural expectation from the inflationary universe. We consider
Friedmann's equation with non-vanishing $\rL$, which provides Hubble's
expansion rate $H=\dot{a}/a$ ($\dot{a}\equiv da/dt$) as a function of the
matter and vacuum energy densities:
\begin{equation}\label{Friedmann}
H^2=\frac{8\pi G}{3}(\rmr+\rL)\,.
\end{equation}
As stated, we assume that $\rL=\rL(t)$ and $G=G(t)$ can be functions of
the cosmic time $t$. We will denote the current value of the Hubble rate
by $H_0\equiv 100$\,\,h$\, Km/s/Mpc$. The observations give $h\simeq
0.70$. The dynamical equation for the acceleration of the universe is:
\begin{equation}\label{acceleration}
\frac{\ddot{a}}{a}=-\frac{4\pi\,G}{3}\,(\rmr+3\pmr-2\rL)=-\frac{4\pi\,G}{3}\,(1+3\wm)\,\rmr+\frac{8\pi\,G}{3}\,\rL\,.
\end{equation}
In the late universe ($\rmr\to 0$) the vacuum energy density $\rL$
dominates. It accelerates the cosmos for $\rL>0$. This may occur
either, because $\rL$ is constant, and for a sufficiently old universe one
finally has $\rmr(t)<2\,\rL$, or because $\rL(t)$ evolves with time, and
the situation $\rL(t)>\rmr(t)/2$ is eventually reached sooner or later
than expected. The general Bianchi identity
$\bigtriangledown^{\mu}G_{\mu\nu}=0$, involving the Einstein tensor on the
\textit{l.h.s.} of Eq.\, (\ref{EE}), leads to the following relation for
the full source tensor on its \textit{r.h.s.} (after we include the CC
term):
\begin{equation}\label{GBI}
\bigtriangledown^{\mu}\left(G\,\tilde{T}_{\mu\nu}\right)=\bigtriangledown^{\mu}\,\left[G\,(T_{\mu\nu}+g_{\mu\nu}\,\rL)\right]=0\,.
\end{equation}
The last equation provides the following ``mixed'' local conservation law:
\begin{equation}\label{BianchiGeneral}
\frac{d}{dt}\,\left[G(\rmr+\rL)\right]+3\,G\,H\,(\rmr+\pmr)=0\,,
\end{equation}
where $G$ and/or $\rL$ may be functions of the cosmic time. Although the
previous equation is not independent of (\ref{Friedmann}) and
(\ref{acceleration}), it is useful to understand the possible transfer of
energy between the vacuum and matter, with or without the participation of
a time-evolving gravitational coupling. For instance, if
{$\dot{\rho}_{\CC}\neq 0$}, matter is not generally conserved, since the
vacuum could decay into matter, or matter could disappear into vacuum
energy (including a possible contribution from a variable $G$, if
$\dot{G}\neq 0$). The local conservation law (\ref{BianchiGeneral}) mixes
the matter-radiation energy density with the vacuum energy $\rL$.

We mention the following possibilities:
\begin{itemize}

\item \textbf{Model I}:  $G=$const. {and} $\rL=$const.:

If there are no other components in the cosmic fluid, this is the
standard case of $\CC$CDM cosmology, implying the local covariant
conservation law of matter-radiation:

\begin{equation}\label{standardconserv}
\dot{\rho}_m+3\,H\,(\rmr+\pmr)=0.
\end{equation}

\item  \textbf{Model II}: $G=$const {and} $\dot{\rho}_{\CC}\neq 0$:

Here Eq.(\ref{BianchiGeneral}) leads to the mixed conservation law:
\begin{equation}\label{mixed conslaw}
\dot{\rho}_{\CC}+\dot{\rho}_m+3\,H\,(\rmr+\pmr)=0\,.
\end{equation}
An exchange of energy between the matter and the vacuum takes place.

\item \textbf{Model III}: $\dot{G}\neq 0$ {and} $\rL=$const.:
\begin{equation}\label{dGneqo}
\dot{G}(\rmr+\rL)+G[\dot{\rho}_m+3H(\rmr+\pmr)]=0\,.
\end{equation}
Since $G$ does not stay constant here, this equation implies a
non-conservation of matter. It could be solved e.g. for $G$,
if $\rmr$ and $\rL$ would be given by some non-conservation
ansatz.

\item \textbf{Model IV}: $\dot{G}\neq 0$ {and} $\dot{\rho}_{\CC}\neq
    0$:

There are many possibilities here. We consider the simplest one by
    assuming the standard local covariant conservation of
    matter-radiation, i.e Eq.\,(\ref{standardconserv}). Eq.\,(\ref{BianchiGeneral})
leads to:
\begin{equation}\label{Bianchi1}
(\rmr+\rL)\dot{G}+G\dot{\rL}=0\,.
\end{equation}
This situation is complementary to the previous one. Here the
dynamical interplay is between $G$ and $\rL$, whereas $\rmr$ is also
time evolving, but decoupled from the feedback between $G$ and $\rL$.

\item \textbf{Model V}: Another possibility with $\dot{G}\neq 0$ {and}
    $\dot{\rho}_{\CC}\neq 0$ is the case that there is no matter in
    the universe: $\rmr=0$. Then Eq.\,(\ref{BianchiGeneral}) implies
    $G\,\rL=$const. This does not exclude that both parameters can be
    time evolving while the product remains constant. This situation
    could only be of interest in the early universe, when matter still
    did not exist and only the vacuum energy was present.

\end{itemize}

Only in the class of Models I and IV matter is covariantly self-conserved,
i.e. matter evolves according to Eq.\,(\ref{standardconserv}). In terms of
the scale factor we find:
\begin{equation}\label{eq:standardconserv2}
\rho'_m(a)+\frac{3}{a}(1+\wm)\,\rmr(a)=0\,.
\end{equation}
The prime indicates $d/da$. Its solution can be expressed as follows:
\begin{equation}\label{solstandardconserv}
\rmr(a)=\rmo\,a^{-3(1+\wm)}=\rmo\,(1+z)^{3(1+\wm)}\,.
\end{equation}
We have expressed the result (\ref{solstandardconserv}) in terms of the
scale factor $a=a(t)$ and the cosmological redshift $z=(1-a)/a$.

We shall focus on Models II, III and IV. Each of these models stands for a
whole class of possible scenarios. One has to introduce more
specifications before being able to perform concrete calculations. The
variation of the ``fundamental constants'' (e.g. $\rL$, $G$) could emerge
as an effective description of some deeper dynamics associated to QFT in
curved space-time, e.g. in quantum gravity or in string theory. This
should provide definite time/redshift-evolution laws $\rL=\rL(z)\,,
G=G(z)$. Examples will be discussed in the next sections.

Other fundamental parameters could also be variable. The fine structure
constant might change in time/redshift -- see e.g.
\,\cite{Uzan11,Chiba11}. However positive evidences \cite{Murphy03} are
questioned \,\cite{No-alphateffect}. The possibility that the fundamental
QCD scale parameter $\LQCD$ of the strong interactions could also be
time-evolving (hence redshift dependent) is of special interest (see
sections \ref{sect:timeMasses} and \,\ref{sect:leakage} for details). This
could lead to the non-conservation of matter in the universe. In this
paper we discuss the possibility that this non-conservation of matter
might be related to the cosmological matter non-conservation. This would
lead to a departure from the standard cosmological scenario.

\mysection{Running vacuum energy and the coupling of gravity}
\label{sect:runningCC}

The running of the vacuum energy and/or the gravitational coupling is
expected in QFT in curved space-time\,\cite{Fossil07,SS09}, see also
\cite{JSP11a} and references therein. Running couplings in flat QFT
provide a useful theoretical tool to investigate theories as QED or QCD.
Here the corresponding gauge coupling constants run with the typical
energy of the process.

In the universe we expect that the running of $\rL$ and $G$ is associated
with the typical energy of the classical gravitational external field
linked to the FLRW metric. Here the Hubble rate $H$ will set the scale,
since it is related to the non-trivial structure of the FLRW background.
The universe in an accelerated expansion ($H\neq 0$, $\dot{H}\neq 0$) is a
space-time with dynamical intrinsic curvature:
\begin{equation}\label{RHHd}
R=-6\left(\,\frac{\ddot{a}}{a}+\frac{\dot{a}^2}{a^2}\,\right)
=-12\,H^2-6\,\dot{H}\,.
\end{equation}
In the effective action of QFT in curved space-time\,\cite{Parker09}
$\rL$ and $G$ should be effective couplings depending on a
mass scale $\mu$. This scale parameterizes the various quantum effects
from the matter fields. In some cases the vacuum energy and the
gravitational coupling can be represented as a power series of $\mu$.
The rates of change are given by:
\begin{eqnarray}\label{seriesRLG}
\frac{d\rL(\mu)}{d\ln\mu^2}&=& \sum_{k=0,1,2,...}\,A_{2k}\,\mu^{2k}=A_0+A_2\,\mu^2+A_4\,\mu^4+...\,,\label{seriesRGL}\\
\frac{d}{d\ln\mu^2}\left(\frac{1}{G(\mu)}\right)&=&
\sum_{k=0,1,2,...}\,B_{2k}\,\mu^{2k}=
B_0+B_2\,\mu^2+B_4\,\mu^4+...\label{seriesRGG}\,.
\end{eqnarray}
Such a ``running'' of $\rL$ and $G$ with $\mu$ reflects the dependence of
the leading quantum effects on a cosmological quantity $\xi$ associated
with $\mu$, hence $\rL=\rL(\xi)$ and $G=G(\xi)$. In cosmology we expect
that the physical scale $\xi$ could be the Hubble rate $H(t)$, or the
scale factor $a(t)$\,\cite{Fossil07}, which in most of the cosmological
past also maps out the evolution of the energy densities with $H$. We will
concentrate here on the setting $\mu=H$, which naturally points to the
non-trivial curvature of the background -- Eq\,(\ref{RHHd}) -- and also to
the typical energy of the FLRW ``gravitons'' attached to the quantum
matter loops contributing to the running of $\rL$ and $G^{-1}$ in a
semi-classical description of gravity. The coefficients $A_{2k}, B_{2k}$
receive contributions from boson and fermion matter fields of different
masses $M_i$. The series (\ref{seriesRLG}) becomes an expansion in powers
of the small quantities $H/M_i$ (see Eq.\,(\ref{seriesLambda}) below).
Only even powers of $H$ are involved, due to the general covariance of the
effective action\,\cite{Fossil07,SS09}\,\footnote{In practice, if one
tries to fit the data with a time dependent CC term which is linear in the
expansion rate, i.e. of the form $\CC\propto H$, the results deviate
significantly from the standard $\CC$CDM predictions\,\cite{BPS09}.}.
These expansions converge very fast for $\mu=H$, since $H/M_i\ll 1$ for
any ordinary particle mass. No other $H^{2n}$-terms beyond $H^2$ (not even
$H^4$) can contribute significantly on the \textit{r.h.s.} of equation
(\ref{seriesRLG}) at any stage of the cosmological history below the GUT
scale $M_X\lesssim M_P$. We find:
\begin{equation}\label{seriesLambda}
\frac{d\rL(\mu)}{d\ln\mu^2}=\frac{1}{(4\pi)^2}\left[\sum_{i}\,c_{i}M_{i}^{2}\,\mu^{2}
+\sum_{i}
\,c'_{i}\,\mu^{4}+\sum_{i}\frac{\,c''_{i}}{M_{i}^{2}}\,\mu^{6}\,\,+...\right]
\equiv \,n_2\,\mu^2+{\cal O}(\mu^4)\,.
\end{equation}
We have omitted the $A_0$ term - it would be of order $M_i^4$. This would
produce a too fast running of $\rL$. This can also be derived from the
fact that all known particles satisfy $\mu<M_i$ (for $\mu=H$). None of
them is an active degree of freedom for the running of $\rL$, and only the
subleading terms are available. Approximately we obtain a simple
expression:
\begin{equation}\label{GeneralPS}
\rL(H)=n_{0}+n_{2}H^{2}\,.
\end{equation}
In view of the boundary condition $\rL(H_0)=\rLo$  it is convenient to
rewrite the coefficients of (\ref{GeneralPS}):
\begin{equation}\label{n0n2}
n_0=\rLo-\frac{3\nu}{8\pi}\,M_P^2\,H_0^2\,,\ \ \ \ \
n_2=\frac{3\nu}{8\pi}\,M_P^2\,.
\end{equation}
We have defined the dimensionless parameter
\begin{equation}\label{nu1}
\nu=\frac{1}{6\pi}\, \sum_{i=f,b} c_i\frac{M_i^2}{M_P^2}\,.
\end{equation}
The sum runs over fermions ($f$) and bosons ($b$) contributing to
the loop. The parameter $\nu$ provides the main coefficient of the
one-loop $\beta$-function for the running of the vacuum energy. The
generic expression (\ref{nu1}) adopts a concrete
form with coefficients $c_i$, depending on the
effective action of the underlying QFT (see e.g. \cite{Fossil07}).
The parameter $\nu$ can have any sign $\sigma=\pm$, depending on whether bosons
or fermions dominate.

It is convenient to write (\ref{nu1}) as follows:
\begin{equation}\label{eq:nu}
\nu=\frac{\sigma}{6\pi}\,\frac{M^2}{M_P^2}\,.
\end{equation}
Here $M^2=|\sum_{i=f,b} c_i{M_i^2}|$ is an effective mass squared
representing all the particles contributing to the running after counting
their multiplicities. For $M=M_P$ we have $|\nu|={\cal O}(10^{-2})$. In
general we expect that the set of $M_i$ includes masses of some GUT theory
with a mass scale $M_X\sim 10^{16}$ GeV ($M\simeq M_X<M_P$). A natural
estimate is in the range $\nu=10^{-6}-10^{-3}$ \,\cite{Fossil07}.

If we would instead take the string scale as the characteristic GUT
scale\,\cite{Cmunoz,stringscale}, then $M/M_P\sim 10^{-2}$, and  $|\nu|$
could move to the upper range $10^{-3}$. For $\nu=0$ we have
$n_2=0$ in Eq.\,(\ref{GeneralPS}). In this case the vacuum energy
remains strictly constant at all times: $\rL=\rLo$, and we recover the
standard situacion of the $\CC$CDM model. For non-vanishing $\nu$  the
evolution law (\ref{GeneralPS} leads to:
\begin{equation}\label{RGlaw2}
 \rL(H)=\rLo+ \frac{3\nu}{8\pi}\,M_P^2\,(H^{2}-H_0^2)\,.
\end{equation}
The expansions (\ref{seriesRLG})-(\ref{seriesRGG}) are correlated by the
Bianchi identity (\ref{BianchiGeneral}). If $\mu=\mu(t)$ is a well defined
invertible function, $d\mu/dt\neq 0$  -- as it is in the case with
$\mu=H(t)$ -- we must have
\begin{equation}\label{BianchiGeneralmu}
\frac{dG}{d\mu}\,\left(\rmr+\rL\right)+G\,\frac{d\rL}{d\mu}+
G\left[\frac{d\rmr}{d\mu}+\frac{3}{a}\,\left(\rmr+\pmr\right)\,\frac{da}{d\mu}\right]=0\,.
\end{equation}
This expression shows, that the dynamical dependence of $\rL$ and $G$  may
not be in the cosmic time $t$ (as in many phenomenological models in the
literature\,\cite{CCvariable}), but in $\mu$. There is a possible
connection of the evolution of $\rL$ with the quantum effects of QFT in a
curved background, i.e. with the running $\rL(\mu)$ in an
expanding universe\,\cite{JSP11a}. Since the quantum effects on $G$ and
$\rL$ must satisfy the above differential constraint, they must be
correlated. If we assume that $\rL$ evolves as indicated in
(\ref{RGlaw2}), the corresponding running of $G$ must fulfill
(\ref{BianchiGeneralmu}). But this is still not enough to determine
$G=G(H)$ explicitly, since it depends on whether matter is conserved or
not. Then one has to have a specific ansatz for the matter
non-conservation equation.

We consider two possibilities. We assume that matter is conserved, as in
Model IV of the previous section. The term
 in brackets on Eq.\,(\ref{BianchiGeneralmu}) vanishes -- see
(\ref{eq:standardconserv2}). Using Friedmann's equation (\ref{Friedmann})
and (\ref{RGlaw2}), we are left with:
\begin{equation}\label{eq:diffG}
\frac{3\,H^2}{8\pi G}\,\frac{dG}{dH}+G\,\frac{3\nu}{4\pi}\,M_P^2
H=0\,.
\end{equation}
After integration we obtain:
\begin{equation}\label{GH}
G(H)=\frac{G_0}{1+\nu\,\ln\left(H^2/H_0^2\right)}\,.
\end{equation}
Here we have defined $G_0=1/M_P^2$, the current value of $G$, i.e.
$G_0=G(H_0)$. From (\ref{GH}) we find:
\begin{equation}\label{diffinvG}
\frac{d}{d\ln H^2}\,\frac{1}{G}=\nu\,M_P^2\,.
\end{equation}
Thus (\ref{GH}) is the solution of (\ref{seriesRGG}), when we
take only the leading term in the expansion, which does not
depend on $\mu=H$. This is consistent, since $1/G$ is a large
quantity and must be dominated by this term. The quantity $\rL$, which in
contrast is a much smaller quantity, can not be dominated by $A_0\sim
M_i^4$, but rather by the next-to-leading term, which is proportional to
$H^2$. The leading term in each case dominates the
corresponding running equation. Higher order corrections (involving more
powers of $H$) are possible, but they are negligible in view of
the current value of $H$.

We mention another simple case, where matter is not conserved. We write
$dG/d\mu=G'(a)\,d\mu/da$, $d\rL/d\mu=\rL'(a)\,d\mu/da$, and
$d\rmr/d\mu=\rmr'(a)\,d\mu/da$. Assuming that $\mu=\mu(a)$ is a well
defined invertible function (which is indeed the case, when $\mu=H$) we
have $d\mu/da\neq 0$. If $G$ is constant, Eq.\,(\ref{BianchiGeneralmu}) simplifies again:
\begin{equation}\label{Bronstein2}
\rho'_\CC(a)+\rho'_m(a)+\frac{3}{a}(1+\wm)\,\rmr(a)=0\,.
\end{equation}
This result is consistent with (\ref{mixed conslaw}). The running of the vacuum
energy is due to the non-conservation of
matter. The solution is well-known (see \cite{JSP11a} and references
therein). The corresponding matter non-conservation law is:
\begin{equation}\label{mRG}
\rmr(a) =\rmo\,a^{-3(1+\wm)(1-\nu)}\,.
\end{equation}
The associated running of the vacuum energy density as a function of the
scale factor is given by:
\begin{equation}\label{CRG}
\rL(a)=\rLo+\frac{\nu\,\rmr^0}{1-\nu}\,\left[a^{-3(1+\wm)(1-\nu)}-1\right]\,.
\end{equation}
The equations (\ref{mRG}) and (\ref{CRG}) do satisfy (\ref{Bronstein2})
and the boundary conditions $\rmr(a=1)=\rmo$ and $\rL(a=1)=\rLo$ are
fulfilled for the present universe. The running vacuum law (\ref{CRG}) is
a consequence of the original equation (\ref{RGlaw2}). The consistency of
these two formulae implies that the invertible function $\mu=\mu(a)$ (i.e.
$H=H(a)$) is given by:
\begin{equation}
{H^2(a)}=\frac{8\pi\,G}{3\,(1-\nu)} \left[{\rLo-\nu\,\rco}+\rmo\,
a^{-3(1+\wm)(1-\nu)}\right] \;. \label{nomalHflow}
\end{equation}
Here $\rco$ is the present value of the critical density. For $\nu=0$ we
obtain a dilution law for the matter density of the form
(\ref{solstandardconserv}), i.e. $\sim a^{-3}$ for non-relativistic and
$\sim a^{-4}$ for relativistic matter; also a constant $\rL=\rLo$, and the
canonical form for the Hubble expansion rate $H=H(a)$.

These  models are compatible with the observational data, both on the
Hubble expansion (e.g. from SNIa+BAO) as well as on structure formation
(power spectrum, growth factor and CMB) for values of the relevant
parameter (\ref{nu1}) up to $|\nu|\sim 10^{-3}$ -- see
\,\cite{BPS09,GSBP11} for details, and \cite{BPS10} for some astrophysical
applications. We shall come back to this cosmological input in
sect.\,\ref{sect:leakage}.

We note the coincidence of this order of magnitude estimate for $\nu$ with
its theoretical expectations for being a $\beta$-function coefficient of
$\rL$. The generalizations of these running vacuum models are possible at
a similar level of phenomenological success, see e.g. \cite{BSP12a}. As
also shown in this reference, alternative dynamical models (such as the
so-called entropic-force cosmologies) are not successful, although they
have many elements in common. There exist time evolving vacuum models,
which can help to cure the old cosmological constant problem and the
coincidence problem\,\cite{RelaxTH,RelaxObsvandLXCDM}.

Thus there exists an interesting class of cosmological models with time
evolving vacuum energy which are phenomenologically acceptable, but not
every phenomenological model can be successfully tested (in this respect
we have also mentioned the unsuccessful cosmologies with vacuum energy
linear in $H$ -- see\,\cite{BPS09} and references therein).

\mysection{Time evolving masses in the Standard Model of particle physics}
\label{sect:timeMasses}

In this section we discuss experiments on the time variation of the
fundamental constants of Nature. We suggest that they could be related to
matter non-conservation.

\mysubsection{The Oklo phenomenon}\label{sect:Oklo}

There are experiments which suggest that the fine structure constant
$\alpha_{\rm em}$ has not remained constant throughout the cosmic
evolution. There are many independent observations suggesting this
possibility\,\cite{Uzan11,Chiba11}. We also mention the ``Oklo
phenomenon''\,\cite{Shlyakhter1976,DamourDyson96,Fujii2000}. It is related
to the natural fission reactor (the Oklo uranium mine) in Gabon (West
Africa), first discovered in 1972 by the French Commissariat \`a
l'\'Energie Atomique. This natural reactor operated nearly 2 billion years
ago for a period of some two hundred thousand years at a power of $\sim
100$ Kw. The data correspond to a process that occurred at the redshift
$z\simeq 0.16$ (for the typical values $h\simeq 0.70$, $\OMo\simeq 0.27$,
$\OLo\simeq 0.73$ of the cosmological parameters). This is the redshift at
which we may be sensitive to variations of the fundamental constants.

The fraction of $^{235}U$ in the Oklo site has decreased since then from
$3.68\%$ to $0.72\%$. This depletion with respect to the current standard
value is a proof of the past existence of a
spontaneous chain reaction. Water from river Oklo provided the moderator
for the neutrons. One of the nuclear fission products is the Samarium's
isotope $^{149}{\rm Sm}_{62}$ which upon neutron capture becomes the
excited isotope of the same element $^{150}{\rm Sm}_{62}$:
\begin{equation}\label{Samarium}
^{149}{\rm Sm}_{62}+n\to\ ^{150}{\rm Sm}_{62}+\gamma\,.
\end{equation}
The sustained fission chain at the Oklo mine leads to the process (\ref{Samarium}).
The relatively light
isotope $^{149}{\rm Sm}_{62}$ is not a fission product of the $^{235}U$,
so the reaction (\ref{Samarium}) took place in the natural ores of Oklo.
It was observed that the ratio of isotopes  $^{149}{\rm
Sm}_{62}/^{147}{\rm Sm}_{62}$ in  samples of Samarium in these ores is
$0.02$, while in normal Samarium is $0.9$. The depletion shows
that the reaction (\ref{Samarium}) took place for a long
time in the Oklo reactor.

The cross section of the neutron capture (\ref{Samarium}) depends on the
energy of a resonance at $E_r = 97.3$ meV and is well described by the
Breit-Wigner formula:
\begin{equation}\label{eq:BreitWigner}
\sigma(E)=\frac{g\pi\hbar^2}{2m_n
E}\,\frac{\Gamma_n\Gamma_{\gamma}}{(E-E_r)^2+\Gamma^2/4}\,.
\end{equation}
Here $g=9/16$ is a spin-dependent statistical factor, $\Gamma$ is the
total width, i.e. the sum of the neutron partial width ($\Gamma_n=0.533$
meV) and of the radiative partial width ($\Gamma_{\gamma}=60.5$ meV). In
order to estimate the cross-section in a more realistic way, one has to
thermal average the above Breit-Wigner formula, using the geophysical
conditions at the Oklo site. From here one can infer the uncertainty in
the resonance energy, $\delta E_r$, which is set equal to $E^{\rm
Oklo}-E_r^0$, where $E^{\rm Oklo}$ is the value of the resonance during
the Oklo phenomenon and $E_r^0$ is the possibly different value taken
today. From the mass formula of heavy nuclei the change in resonance
energy is related to $\alpha_{\rm em}$ through the Coulomb energy
contribution:
\begin{equation}\label{eq:Coulomb}
\delta E_r=-1.1\,\frac{\delta\alpha_{\rm em}}{\alpha_{\rm em}}\,{\rm
MeV}\,.
\end{equation}
From the estimates on $\delta E_r$ (ranging from a dozen meV to a hundred
MeV\,\cite{DamourDyson96,Fujii2000}) one infers from (\ref{eq:Coulomb}) a
tight bound on the time variation of the fine structure constant of order
$\dot{\alpha}_{\rm em}/\alpha_{\rm em}\sim 10^{-17}\,{\rm yr}^{-1}\,$.
This is comparable to the best bounds from atomic
clocks\,\cite{Uzan11,Chiba11}. But the debate continues on the reliability
of the data obtained in the Oklo mine. Even if the corresponding bound,
obtained on the time variation of the electromagnetic coupling, is
eventually validated, the Oklo phenomenon cannot easily provide
information on the time variation of the strength of the nuclear
interaction, since it is sensitive only to dimensionless ratios of nuclear
quantities. It cannot be used to extract a possible variation of the QCD
scale parameter $\LQCD$. This is essential to establish a link between the
time variation of fundamental nuclear and particle physics constants with
the the corresponding variation of the vacuum energy density in the the
cosmic expansion.

\mysubsection{Time variation of the fundamental QCD constant:
implications for the nucleon mass and the nuclear masses in the universe
}\label{sect:AtomicMasses}

It has been argued that the fundamental QCD scale parameter $\LQCD$ could
vary much faster than $\alpha_{\rm
em}$\,\cite{Fritzsch11,Fritzsch06,Langacker02}. This change would be
related to a corresponding change of the nucleon mass. Within the context
of QCD the nucleon mass and the other hadronic masses are determined by
the value of the QCD scale parameter $\LQCD$. The leading contribution to
the nucleon mass can be expressed as $m_N\simeq c_{\rm QCD}\LQCD$, where
$c_{\rm QCD}$ is a non-perturbative coefficient. The masses of the light
quarks $m_u$, $m_d$ and $m_s$ also contribute to the the proton mass,
although by less than $10\%$ only. There is also a small contribution from
electromagnetism. Let us take for instance the proton mass $m_p\simeq 938$
MeV. It can be computed from the QCD scale parameter $\LQCD$, the quarks
masses and the electromagnetic contribution:
\begin{eqnarray}\label{eq:ProtonMass}
m_p &=&c_{\rm QCD}\LQCD+c_u\,m_u+c_d\,m_d+c_s\,m_s+c_{\rm em}\LQCD\nonumber\\
&=&\left(860+21+19+36+2\right)\,{\rm MeV}\,.
\end{eqnarray}
The QCD scale parameter is related to the strong coupling constant
$\alpha_s=g_s^2/(4\pi)$. To lowest (1-loop) order one finds:
\begin{equation}\label{alphasLQCD}
\alpha_s(\mu_R)=\frac{1}{\beta_0\,\ln{\left(\LQCD^2/\mu_R^2\right)}}=\frac{4\pi}{\left(11-2\,n_f/3\right)\,\ln{\left(\mu_R^2/\LQCD^2\right)}}\,,
\end{equation}
where $\mu_R$ is the renormalization point and $\beta_0\equiv-b_0=
-(33-2\,n_f)/(12\,\pi)$ ($n_f$ being the number of quark flavors) is the
lowest order coefficient of the $\beta$-function.

The QCD scale parameter $\LQCD$ has been measured: $\LQCD = 217\pm 25$
MeV. When we embed QCD in the FLRW expanding background, the value of
$\LQCD$ need not remain rigid anymore. The value of $\LQCD$ could change
with $H$, and this would mean a change in the cosmic time. If
$\LQCD=\LQCD(H)$ is a function of $H$, the coupling constant
$\alpha_s=\alpha_s(\mu_R;H)$ is also a function of $H$ (apart from a
function of $\mu_R$). The relative cosmic variations of the two QCD
quantities are related (at one-loop) by:
\begin{equation}\label{eq:timealphaLQC}
\frac{1}{\alpha_s}\frac{d\alpha_s(\mu_R;H)}{dH}=\frac{1}{\ln{\left(\mu_R/\LQCD\right)}}\,\left[\frac{1}{\LQCD}\,\frac{d{\Lambda}_{\rm
QCD}(H)}{dH}\right]\,.
\end{equation}
If the QCD coupling constant $\alpha_s$ or the QCD scale
parameter $\LQCD$ undergo a small cosmological time shift, the nucleon
mass and the masses of the atomic nuclei would also change in proportion
to $\LQCD$.

The cosmic dependence of the strong coupling $\alpha_s(\mu_R;H)$ can be
generalized to the other couplings
$\alpha_i=\alpha_i(\mu_R;H)$\,\cite{Fritzsch06}. In a grand unified theory
these couplings converge at the unification point. Let $d\alpha_i$ be the
cosmic variation of $\alpha_i$ with $H$. Each of the $\alpha_i$ is a
function of $\mu_R$, but the expression
$\alpha_i^{-1}\left(d\alpha_i/\alpha_i\right)$ is independent of $\mu_R$.
One can show that the running of $\alpha_{\rm em}$ is related to the
corresponding cosmic running of $\LQCD$ as follows:
\begin{equation}\label{eq:timealpha}
\frac{1}{\alpha_{\rm em}}\frac{d\alpha_{\rm
em}(\mu_R;H)}{\,dH}=\frac83\,\frac{\alpha_{\rm
em}(\mu_R;H)/\alpha_s(\mu_R;H)}{\ln{\left(\mu_R/\LQCD\right)}}\,\left[\frac{1}{\LQCD}\,\frac{d{\Lambda}_{\rm
QCD}(H)}{dH}\right]\,.
\end{equation}
At the renormalization point $\mu_R=M_Z$, where both $\alpha_{\rm em}$ and
$\alpha_s$ are well-known, one finds:

\begin{equation}\label{eq:timealpha2}
\frac{1}{\alpha_{\rm em}}\frac{d\alpha_{\rm
em}(\mu_R;H)}{\,dH}\simeq
0.03\left[\frac{1}{\LQCD}\,\frac{d{\Lambda}_{\rm
QCD}(H)}{dH}\right]\,.
\end{equation}

Thus the electromagnetic fine structure constant runs more than $30$ times
slower with the cosmic expansion than $\LQCD$. Searching for a cosmic
evolution of $\LQCD$ is much easier than searching for the time
variation of $\alpha_{\rm em}$.

\mysubsection{Time evolution of the proton - electron mass ratio}
\label{sect:timeProtonElectron}

We consider the mass ratio:
\begin{equation}\label{eq:mupe}
\mupe\equiv\frac{m_p}{m_e}\,.
\end{equation}\\
This ratio is known with high accuracy:
$\mupe=1836.15267247(80)$\,\cite{Mohr2008}. Since a change of $\LQCD$
would not affect the electron mass, the mass ratio (\ref{eq:mupe}) would
change during the cosmological evolution.

First we consider astrophysical tests. The spectrum of
$H_2$ provides a direct
operational handle to test possible variations of (\ref{eq:mupe}).
Particularly significant is the study of Ref.\cite{Reinhold06}, based on
comparing the $H_2$ spectral Lyman and Werner lines, observed
in the Q 0347-383 and Q 0405-443 quasar absorption systems, with the
laboratory measurements.

The result indicates, that $\mupe$ could have decreased in the past $12$
Gyr, corresponding to a relative time variation of
\begin{equation}\label{eq:timemupe}
\frac{\dot{\mu}_{pe}}{\mupe}=(-2.16\pm 0.52)\times 10^{-15}\,{\rm
yr}^{-1}\,.
\end{equation}
It has been pointed more recently by other authors\,\cite{King2008} that
this measurement may suffer from spectral wavelength calibration
uncertainties, and the reanalysis of the time variation would show a
significance at the $1\,\sigma$ level only.

Now we consider laboratory tests, using atomic clocks. According to our
estimate (\ref{eq:timealpha2}), the largest effect is expected to be a
cosmological {redshift} ({hence time variation}) of the nucleon mass,
which can be observed by monitoring molecular frequencies. These are
precise experiments in quantum optics, e.g. obtained by comparing a cesium
clock with 1S-2S hydrogen transitions. In a cesium clock the time is
measured by using a hyperfine transition\,\footnote{Recall that the cesium
hyperfine clock provides the modern definition of time. In SI units, the
second is defined to be the duration of $9.192631770\times 10^{9}$ periods
of the transition between the two hyperfine levels of the ground state of
the $^{133}$Cs atom}. Since the frequency of the clock depends on
the magnetic moment of the cesium nucleus, a possible variation of the
latter is proportional to a possible variation of $\LQCD$. A hyperfine
splitting is a function of $Z\,\alpha_{\rm em}$ ($Z$ being the atomic
number) and is proportional to $Z\,\alpha_{\rm
em}^2(\mu_N/\mu_B)(m_e/m_p)\,R_{\infty}$, where $R_{\infty}$ is the
Rydberg constant, $\mu_N$ is the nuclear magnetic moment and
$\mu_B=e\hbar/2m_pc$ is the nuclear magneton. We have
$\dot{\mu}_N/\mu_N\propto -\dot{\Lambda}_{\rm QCD}/\LQCD$. The hydrogen
transitions are only dependent on the electron mass, which
we assume to be constant. The comparison over a period of time between the
cesium clock with hydrogen transitions provides an atomic laboratory
measurement of the ratio (\ref{eq:mupe}). The most recent atomic clock
experiment at the MPQ (Max-Planck-Institut f\"ur Quantenoptik) at Garching
near Munich gives a limit\,\cite{Haensch12}:
\begin{equation}\label{eq:timeLambda}
\left|\frac{\dot{\Lambda}_{\rm QCD}}{\LQCD}\right|<10^{-14}\,{\rm
yr}^{-1}\,.
\end{equation}
Since the proton mass is given essentially by $\LQCD$, as indicated by
Eq.\,(\ref{eq:ProtonMass}), we have $\dot{m}_p\simeq
c_{\LQCD}\,\dot{\Lambda}_{\rm QCD}$. The corresponding time variation of
the ratio (\ref{eq:mupe}) would be:
\begin{equation}\label{eq:timeLambda2}
\left|\frac{\dot{\mu}_{pe}}{\mupe}\right|=\left|\frac{\dot{m}_p}{m_p}\right|\simeq
\left|\frac{\dot{\Lambda}_{\rm QCD}}{\LQCD}\right|<
10^{-14}\,{\rm yr}^{-1}\,.
\end{equation}
Thus the atomic clock result (\ref{eq:timeLambda}) would indicate a time
variation of the ratio $\mupe$, which is consistent (in absolute value)
with the astrophysical measurement (\ref{eq:timemupe}). The result above
implies also a bound for a possible time variation of the light quark masses:

\begin{equation}\label{eq:timemq}
\left|\frac{\dot{m}_{q}}{m_q}\right|\lesssim 10^{-14}\,{\rm
yr}^{-1}\,.
\end{equation}
%


\mysection{Dynamical dark energy and a
cosmic link with nuclear and particle physics }\label{sect:leakage}

The time evolution of the fundamental ``constants'' $\rL$ and $G$ of
gravity could be related to the time variation of the fundamental
``constants'' in nuclear and particle physics. In some models one can have
matter conservation even though $\rL$ is running, but at the expense of
having a running $G$ as well -- confer Model IV of sect. \ref{sect:timeCC}
and Eq.\,(\ref{Bianchi1}). In an alternative class of models, $G$ runs
thanks to the non-conservation of matter, as in Model III of sect.
\ref{sect:timeCC}, but then $\rL$ stays fixed. If $G$ stays fixed and
$\rL$ is evolving, there is a transfer of energy from matter into the
vacuum, or vice versa -- cf. the Model II class of sect. \ref{sect:timeCC}
and Eq.\,(\ref{mixed conslaw}). The various classes of cosmological
scenarios are interesting, but the last two could help us to understand
the potential cosmic time variation of the fundamental ``constants'' of
nuclear and particle physics, such as the QCD scale, the nucleon mass and
the masses of nuclei.

\mysubsection{Non-conservation of matter at fixed $G$}
\label{sect:leakage1}

First we consider the class of scenarios denoted as Model II. Let $\rMo$
be the total matter density of the present universe, which is essentially
non-relativistic ($\wm\simeq 0$). The corresponding normalized density is
$\OMo=\rMo/\rco\simeq 0.27$, where $\rco$ is the current critical density.
Similarly, $\OLo=\rLo/\rco\simeq 0.73$ is the current normalized vacuum
energy density, for flat space. If $\rL$ evolves with the Hubble rate in
the form indicated in Eq.\,(\ref{RGlaw2}), the non-relativistic matter
density and vacuum energy density evolve with the scale factor, given in
(\ref{mRG}) and (\ref{CRG}). Expressing the result in terms of the
cosmological redshift $z=(1-a)/a$, we find:
\begin{equation}\label{mRG2}
\rM(z;\nu) =\rMo\,(1+z)^{3(1-\nu)}\,,
\end{equation}
and
\begin{equation}\label{CRG2}
\rL(z)=\rLo+\frac{\nu\,\rM^0}{1-\nu}\,\left[(1+z)^{3(1-\nu)}-1\right]\,.
\end{equation}
The crucial parameter is $\nu$, which we have introduced in
sect.\,\ref{sect:runningCC}. It is responsible for the time evolution of
the vacuum energy. From Eq.\,(\ref{mRG2}) we confirm, that it accounts
also for the non-conservation of matter, since it leads to the exact local
covariant conservation law (\ref{eq:standardconserv2}). For
non-relativistic matter we find:
\begin{equation}\label{mRG2b}
\rM(z) =\rMo\,(1+z)^{3}\,.
\end{equation}
$\delta\rho_M\equiv \rM(z;\nu)-\rM(z)$ is the net amount of
non-conservation of matter per unit volume at a given redshift. This
expression must be proportional to $\nu$, since we subtract the conserved
part. At this order we have $\delta\rho_M=-3\,\nu\, \rMo (1+z)^3\ln(1+z)$.
{We differentiate it} with respect to time and expand in $\nu$, and
finally divide the final result by $\rM$. {This provides} the relative
time variation:
\begin{equation}\label{eq:deltadotrho}
\frac{\delta\dot{\rho}_M}{\rM}=3\nu\,\left(1+3\ln(1+z)\right)\,H+{\cal O}(\nu^2)\,.
\end{equation}
Here we have used $\dot{z}=(dz/da)\dot{a}=(dz/da)aH=-(1+z)H$. Assuming
relatively small values of the redshift, we may neglect the log term and
are left with:
\begin{equation}\label{eq:deltadotrho2}
 \frac{\delta\dot{\rho}_M}{\rM}\simeq 3\nu\,\,H\,.
\end{equation}
From (\ref{CRG2}) we find:
\begin{equation}\label{eq:deltaLambda}
\frac{\dot{\rho}_{\CC}}{\rL}\simeq -3\nu\,\frac{\OMo}{\OLo}\,(1+z)^3\,H+{\cal O}(\nu^2)\,.
\end{equation}
It is of the same order of magnitude as (\ref{eq:deltadotrho2}) and has
the opposite sign. Let us compare the theoretical expression
(\ref{eq:deltadotrho2}) with the experimental results (\ref{eq:timemupe})
and (\ref{eq:timeLambda2}), described in the previous section. Taking the
current value of the Hubble parameter as a reference, $H_0=1.0227\,h\times
10^{-10}\,{\rm yr}^{-1}\,$, where $h\simeq 0.70$, we obtain $|\nu|\lesssim
{\cal O}(10^{-4})$ for the most conservative case. It is a rather tight
bound, in accordance with the QFT expectations in
sect.\,\ref{sect:runningCC}.

What is the role played by the running vacuum energy (\ref{CRG2})? Its
evolution in combination with the non-conservation of matter affects many
relevant cosmological observables, which have currently been measured with
high precision. From a detailed analysis of the combined data on type Ia
supernovae, the Cosmic Microwave Background (CMB), the Baryonic Acoustic
Oscillations (BAO) and the structure formation data a direct cosmological
bound on $\nu$ has been obtained in the literature\,\,\cite{BPS09,GSBP11}:

\begin{equation}\label{eq:numodeliicosmology}
|\nu|^{\rm cosm.}\lesssim {\cal O}(10^{-3})\,,\ \ \ \ \ ({\rm Model\ II\ sect.\ \ref{sect:timeCC}})\,.
\end{equation}
It is consistent with the theoretical expectations.
In the next section we analyze another model which can also accommodate
matter non-conservation in the form (\ref{mRG2}), but at the expense of a
time varying $G$. We compare it with a similar model, where matter is
conserved.

\mysubsection{Non-conservation of matter at fixed $\rL$}
\label{sect:leakage2}

Within the class of scenarios indicated as Model III of
sect.\,\ref{sect:runningCC} the parameter $\rL$ remains constant
($\rL=\rLo$) and $G$ is variable. This is possible due to the presence of
the \textit{non} self-conserved matter density (\ref{mRG2}). Trading the
time variable by the scale factor, we can rewrite Eq.\,(\ref{dGneqo}) as
follows:
\begin{equation}\label{dGneqo2}
G'(a)\left[\rM(a)+\rLo\right]+G(a)\left[{\rho}'_M(a)+\frac{3}{a}\,\rM(a)\right]=0\,.
\end{equation}
The primes indicate differentiation with respect to the scale factor. We
insert equation (\ref{mRG2}) in (\ref{dGneqo2}), integrate the resulting
differential equation for $G(a)$ and express the final result in terms of
the redshift:
\begin{equation}\label{GzfixedL}
G(z)=G_0\,\left[\OMo\left(1+z\right)^{3(1-\nu)}+\OLo\right]^{\nu/(1-\nu)}\,.
\end{equation}
Here $G_0=1/M_P^2$ is the current value of the gravitational coupling. The
previous equation is correctly normalized: $G(z=0)=G_0$, due to the cosmic
sum rule in flat space: $\OMo+\OLo=1$. For $\nu=0$ the gravitational
coupling $G$ remains constant: $G=G_0$. {Since $\rL$ is constant in the
current scenario}, the small variation of $G$ is entirely due to the
non-vanishing value of the $\nu$-parameter in the matter non-conservation
law (\ref{mRG2}). This leads to the dynamical feedback of $G$ with
matter\,\footnote{This feedback can also be conceived in the context of
gravitation holography\,\cite{Guberina2006} if one also takes as a
starting point the matter non-conservation law (\ref{mRG2}). This law was
first suggested and analyzed in \cite{oldCCstuff1} and later on in
\cite{scalingmatter2}.}. For the present model Friedmann's equation
(\ref{Friedmann}) becomes:
\begin{equation}\label{FriedmannModeliii}
H^2(z)=\frac{8\pi
G(z)}{3}\left[\rMo\,(1+z)^{3(1-\nu)}+\rLo\right]=H_0^2\frac{G(z)}{G_0}\,\left[\OMo\,(1+z)^{3(1-\nu)}+\OLo\right].
\end{equation}
Combining (\ref{GzfixedL}) and (\ref{FriedmannModeliii}), we find the
Hubble function of this model in terms of $z$:
\begin{equation}\label{HzModeliii}
H^2(z)=H_0^2\,\left[\OMo\left(1+z\right)^{3(1-\nu)}+\OLo\right]^{1/(1-\nu)}\,,
\end{equation}
and we obtain:
\begin{equation}\label{GrelatedH}
\frac{G(z)}{G_0}=\left[\frac{H^2(z)}{H_0^2}\right]^{\nu}\,.
\end{equation}
Since $\nu$ is presumably  small in absolute value (as in the previous
section), we can expand (\ref{GrelatedH}) in this parameter:
\begin{equation}\label{GzfixedLsmallnu}
G(H)\simeq G_0\,\left(1+\nu\,\ln\frac{H^2}{H_0^2}+{\cal
O}(\nu^2)\right)\,.
\end{equation}
At leading order in  $\nu$ this expression for the variation of $G$ is
identical to the one found for Model IV of sect.\,\ref{sect:timeCC}, see
Eq.\,(\ref{GH}), except for the sign of $\nu$. The
equation(\ref{GzfixedLsmallnu}) allows us to estimate the value of the
parameter $\nu$ by confronting the model with the experimental data on the
time variation of $G$. Differentiating (\ref{GzfixedLsmallnu}) with
respect to the cosmic time, we find in leading order in $\nu$:
\begin{equation}\label{GdotoverGo}
\frac{\dot{G}}{G}= 2\nu\,\frac{\dot{H}}{H}=-2\,(1+q)\,\nu\,H\,,
\end{equation}
where we have used the relation $\dot{H}=-(1+q)H^2$, in which
$q=-\ddot{a}/(aH^2)$ is the deceleration parameter.  From the known data
on the relative time variation of $G$ the bounds indicate that
$|\dot{G}/G|\lesssim 10^{-12}\,{\rm yr}^{-1}\,$\,\cite{Uzan11,Chiba11}. If
we take the present value of the deceleration parameter, we have
$q_0=3\OMo/2-1=-0.595\simeq -0.6$ for a flat universe with $\OMo=0.27$. It
follows:
\begin{equation}\label{GdotoverGo2}
\left|\frac{\dot{G}}{G_0}\right|\lesssim \,0.8 |\nu|\,H\,.
\end{equation}
Taking the current value of the Hubble parameter: $H_0\simeq 7\times
10^{-11}\,{\rm yr}^{-1}\,$ (for $h\simeq 0.70$), {we obtain $|\nu|\lesssim
10^{-2}$. The real value of $|\nu|$ can be smaller, but to compare the
upper bound that we have obtained with observations makes sense in view of
the usual interpretation of $\nu$ in sect.\,\ref{sect:runningCC} and the
theoretical estimates indicated there. The constraints from Big Bang
nucleosynthesis (BBN) for the time variation of $G$ are more stringent and
lead to the improved bound:
\begin{equation}\label{eq:nuboundGvariable}
|\nu|^{\rm BBN}\lesssim 10^{-3}\,,\ \ \ \ \ ({\rm Model\ III\ sect.\ \ref{sect:timeCC}})\,.
\end{equation}
This bound can be obtained by adapting the study of Ref.\,\cite{GSFS10},
which was made for Model IV of sect.\,\ref{sect:timeCC}. Since Models III
and IV share a similar kind of running law for the gravitational coupling
(except for the sign of $\nu$) --  confer equations (\ref{GH}) and
(\ref{GzfixedLsmallnu}) ---, we can extract the same bound for $|\nu|$ in
the two models following the method of sect. 5.2 of Ref.\,\cite{GSFS10}
and references therein, particularly\,\cite{{Iocco:2008va}}. The final
result is Eq.\,(\ref{eq:nuboundGvariable}). The cosmological data from
different sources furnish about the same upper bound on $|\nu|$ for the
two  running models where matter is non-conserved, i.e. Models II and III
of sect.\,\ref{sect:timeCC}. In both cases the upper bound on $|\nu|$ is
$\sim 10^{-3}$, as shown by equations (\ref{eq:numodeliicosmology}) and
(\ref{eq:nuboundGvariable}).

The previous bounds on $|\nu|$ for Models II and III are completely
general (meaning that they apply to all forms of matter), since they are
obtained from cosmological data tracing the possible evolution of $\rL$
and $G$, respectively. But these cosmological bounds are weaker than those
that follow, if we interpret $\nu$ as a matter non-conservation parameter.
Since matter is indeed non-conserved in both of these models,
Eq.\,(\ref{eq:deltadotrho2}) and the lab bound (\ref{eq:timeLambda}) do
apply in the present case, but \textit{only} if the non-conserved matter
is of nuclear nature. In this case we obtain the stronger constraint
\begin{equation}\label{nuboundfromLab}
|\nu|^{\rm lab.}\lesssim {\cal O}(10^{-4})\,,\ \ \ \ ({\rm Models\ II\ and\ III\ \ sect.\ \ref{sect:timeCC}})\,.
\end{equation}
But if the non-conserved matter is dark matter, then only the weaker
(purely cosmological) bound (\ref{eq:nuboundGvariable}) is valid (see the
next section for a detailed discussion on the distinct contributions from
nuclear matter and dark matter).

Despite $|G|$ varies with time in a comparable way in Models III and IV,
the stronger bound (\ref{nuboundfromLab}) does \textit{not} apply for
Model IV, since matter is conserved in it and hence Eq.\,(\ref{mRG2}) does
not hold for this model. Only the pure BBN cosmological bound
(\ref{eq:nuboundGvariable}) is applicable in this case. This primordial
nucleosynthesis bound on Model IV coincides with an independent bound
obtained for this model from type Ia supernovae, the Cosmic Microwave
Background, the Baryonic Acoustic Oscillations and the structure formation
data (cf. \,\cite{GSBP11} for details). For Model IV two independent
cosmological bounds (BBN plus the current cosmological data) converge to
the same result:
\begin{equation}\label{eq:numodelivcosmology}
|\nu|^{\rm BBN+cosm.}\lesssim {\cal O}(10^{-3})\,,\ \ \ \ \ ({\rm Model\ IV\ sect.\ \ref{sect:timeCC}})\,.
\end{equation}
Although the order of magnitude of the bounds on $|\nu|$ are sometimes
coincident for different models, they are different. For example, Model IV
cannot -- in contrast to Models II and III -- be used to explain the
possible time variation of the fundamental constants of the strong
interactions and the particle masses. It can only be used to explain the
time variation of the cosmological parameters $\rL$ and $G$ in a way which
is independent from the microphysical phenomena in particle physics and
nuclear physics.

Finally, we note that the above cosmic changes in the values of the proton
to electron mass ratio and $G$ or $\rL$ can be written in terms of
dimensionless quantities (in natural units). For example, for Model II
(where $G$ is fixed and $\rL$ is variable) we can define the dimensionless
quantity $\lambda\equiv\CC/m_p^2=8\pi\,G\,\rL/m_p^2$. Then,
\begin{equation}\label{eq:GCC}
\frac{1}{\lambda}\frac{d\lambda}{dt}=\frac{\dot{\rho}_{\CC}}{\rL}-2\,\frac{\dot{m}_p}{m_p}\propto\nu\,H\,,
\end{equation}
because both terms on the \textit{r.h.s.} are proportional to $\nu$ (cf.
sect.\,\ref{sect:leakage1}). Similarly, for Model III (where $\rL$ is
fixed and $G$ is variable) we can construct the dimensionless quantity
$G\,m_p^2$. Its relative variation is also proportional to $\nu$:
\begin{equation}\label{eq:Gmp2}
\frac{1}{G\,m_p^2}\frac{d(G\,m_p^2)}{dt}=\frac{\dot{G}}{G}+2\,\frac{\dot{m}_p}{m_p}\propto\nu\,H\,.
\end{equation}

\mysubsection{Non-conservation of baryonic matter versus dark matter and
the cosmic evolution of $\LQCD$} \label{sect:leakage3}

Here we focus on the impact of the cosmological Models II and III of
sect.\,\ref{sect:timeCC} on the non-conservation of matter in the
universe.  In the previous section we have considered bounds on the
``leakage parameter'' $\nu$ within the class of these models based on the
non-conservation matter density law (\ref{mRG2}). We must be careful in
interpreting such a non-conservation law. For example, if we take the
baryonic density in the universe, which is essentially the mass density of
protons, we can write $\rM^B=n_p\,m_p$, where $n_p$ is the number density
of protons and $m_p^0=938.272 013(23)$ MeV is the current proton mass. If
this mass density is non-conserved, either $n_p$ does not exactly follow
the normal dilution law with the expansion, i.e. $n_p\sim a^{-3}=(1+z)^3$,
{but} the anomalous law:
\begin{equation}\label{eq:nonconservednumber}
n_p(z)=n_p^0\,(1+z)^{3(1-\nu)}\ \ \ \ ({\rm at\ fixed\ proton\ mass}\ m_p=m_p^0)\,,
\end{equation}
and/or the proton mass $m_p$ does not stay constant with time and
redshifts with the cosmic evolution:
\begin{equation}\label{eq:nonconservedmp}
m_p(z)=m_p^0\,(1+z)^{-3\nu}\ \ \ \ ({\rm with\ normal\ dilution}\ n_p(z)=n_p^0\,(1+z)^3)\,.
\end{equation}
In all cases it is assumed that the vacuum absorbs the difference (i.e.
$\rL=\rL(z)$ ``runs with the expansion''). The first possibility implies
that during the expansion a certain number of particles (protons in this
case) are lost into the vacuum (if $\nu<0$; or ejected from it, if
$\nu>0$), whereas in the second case the number of particles is strictly
conserved. The number density follows the normal dilution law with the
expansion, but the mass of each particle slightly changes (decreases for
$\nu<0$, or increases for $\nu>0$) with the cosmic evolution.

Here we adopt the second point of view, {i.e.}
Eq.\,(\ref{eq:nonconservedmp}). We can interpret the tight bounds from the
laboratory and cosmological observations summarized in
sect.\,\ref{sect:timeMasses} as direct bounds on the cosmic time evolution
of $\LQCD$ (hence on $m_p$ and on the nuclei in the universe). Since the
contribution of the quark masses $m_u$,$m_d$ and $m_s$ to the proton mass
is small -- cf. Eq.\,(\ref{eq:ProtonMass}) -- we can approximate the
proton mass by $m_p\simeq c_{\rm QCD}\,\LQCD$. It will be sufficient to
take into account the leading effects of the time variation of $m_p$
through the corresponding effects in $\LQCD$.

Since the matter content of the universe is dominated by the dark matter
(DM), we cannot exclude that it also varies with cosmic time. Let us
denote the mass of the dominant DM particle $m_X$, and let  $\rX$ and $n_X$
be its mass density and number density, respectively. The overall matter
density of the universe can be written as follows:
\begin{eqnarray}\label{eq:MdensityUniv}
\rM&=&\rB+\rLep+\rR+\rX=\left(n_p\,m_p+n_n\,m_n\right)+ n_e\,m_e+\rR+n_X\,m_X\nonumber\\
&\simeq& n_p\,m_p+n_n\,m_n+n_X\,m_X\,.
\end{eqnarray}
Here $n_p, n_n, n_e, n_X\, (m_p,m_n,m_e,m_X)$ are the number densities
(and masses) of protons, neutrons, electrons and DM
particles. The baryonic and leptonic parts are $\rB=n_p\,m_p+n_n\,m_n$ and
$\rLep=n_e\,m_e$ respectively. The small ratio $m_e/m_p\simeq 5\times
10^{-4}$ implies that the leptonic contribution to the total mass density
is negligible: $\rLep\ll \rB$. We have also neglected the relativistic
component $\rR$ (photons and neutrinos).

If we assume that the mass change through the cosmic evolution is due to
the time change of $m_p$, $m_n$ and $m_X$, we can compute the mass density
anomaly per unit time, i.e. the deficit or surplus with respect to the
conservation law,  by differentiating (\ref{eq:MdensityUniv}) with respect
to time and subtracting the ordinary (i.e. fixed mass) time dilution of
the number densities. The result is:
\begin{equation}\label{eq:timeMdensityUniv}
\delta\dot{\rho}_M=n_p\,\dot{m}_p+n_n\,\dot{m}_n+n_X\,\dot{m}_X\,.
\end{equation}
The relative time variation of the mass density anomaly can be estimated
as follows:
\begin{equation}\label{eq:reltimeMdensityUniv1}
\frac{\delta\dot{\rho}_M}{\rM}=\frac{n_p\,\dot{m}_p+n_n\,\dot{m}_n+n_X\,\dot{m}_X}{n_p\,m_p+n_p\,m_p+n_X\,m_X}\simeq
\frac{n_p\,\dot{m}_p+n_n\,\dot{m}_n+n_X\,\dot{m}_X}{n_X\,m_X}\,\left(1-\frac{n_p\,m_p+n_n\,m_n}{n_X\,m_X}\right)\,.
\end{equation}
The current normalized DM density $\ODMo=\rX/\rc\simeq 0.23$ is
significantly larger than the corresponding normalized baryon density
$\OMBo=\rB/\rc\simeq 0.04$. Therefore $n_X\,m_X$ is larger than
$n_p\,m_p+n_n\,m_n$  by the same amount. If we assume
$\dot{m}_n=\dot{m}_p$, we find approximately:
\begin{equation}\label{eq:reltimeMdensityUniv2}
\frac{\delta\dot{\rho}_M}{\rM}=\frac{n_p\,\dot{m}_p}{n_X\,m_X}\,\left(1+\frac{n_n}{n_p}-\frac{\OMB}{\ODM}\right)+
\frac{\dot{m}_X}{m_X}\left(1-\frac{\OMB}{\ODM}\right)\,.
\end{equation}
In the approximation $m_n=m_p\,$ we can
rewrite the prefactor on the \textit{r.h.s} of
Eq.\,(\ref{eq:reltimeMdensityUniv2}) as follows:
\begin{equation}\label{eq:prefactor}
\frac{n_p\,\dot{m}_p}{n_X\,m_X}=\frac{\OMB}{\ODM}\,\frac{\dot{m}_p}{m_p}\left(1-\frac{n_n/n_p}{1+n_n/n_p}\right)\simeq
\frac{\OMB}{\ODM}\,\frac{\dot{m}_p}{m_p}\left(1-\frac{n_n}{n_p}\right)\,.
\end{equation}
The ratio $n_n/n_p$ is of order $10\%$
after the primordial nucleosynthesis. Since $\OMB/\ODM$ is also of order
$10\%$, we can neglect the product of this term with $n_n/n_p$\,. When
we insert the previous equation into (\ref{eq:reltimeMdensityUniv2}), the
two $n_n/n_p$ contributions cancel each other. The expression
${1-\OMB}/{\ODM}$ factorizes in the two terms on the \textit{r.h.s} of
Eq.\,(\ref{eq:reltimeMdensityUniv2}). The final result is:
\begin{equation}\label{eq:reltimeMdensityUniv}
\left(1-\frac{\OMB}{\ODM}\right)^{-1}\frac{\delta\dot{\rho}_M}{\rM}=\frac{\OMB}{\ODM}\,\frac{\dot{m}_p}{m_p}+\frac{\dot{m}_X}{m_X}
=\frac{\OMB}{\ODM}\,\frac{\dot{\Lambda}_{\rm QCD}}{\LQCD}+\frac{\dot{m}_X}{m_X}\,.
\end{equation}
We have used $m_p\simeq c_{\rm QCD}\,\LQCD$, the latter being accurate up
to $10\%$ corrections at most -- see (\ref{eq:ProtonMass}). Equation
(\ref{eq:reltimeMdensityUniv}) should be a good approximation (at
most $10\%$ corrections).

The expression ${\delta\dot{\rho}_M}/{\rM}$ in
Eq.\,(\ref{eq:reltimeMdensityUniv}) must be the same as the one we have
computed in (\ref{eq:deltadotrho}), if we consider the models based on the
generic matter non-conservation law (\ref{mRG2}). Therefore the two
expressions should be equal, and we obtain approximately:
\begin{equation}\label{eq:reltimeMdensityUniv3}
3\nueff\,H=\frac{\OMB}{\ODM}\,\frac{\dot{\Lambda}_{\rm QCD}}{\LQCD}+\frac{\dot{m}_X}{m_X}\,,
\end{equation}
where we have defined
\begin{equation}\label{eq:nueff}
\nueff=\frac{\nu}{1-\OMB/\ODM}\,.
\end{equation}
We have $\nueff\simeq 1.2\,\nu$. The differential equation
(\ref{eq:reltimeMdensityUniv3}) describes approximately the connection
between the matter non-conservation law (\ref{mRG2}), the evolution of the
vacuum energy density $\rL$ (and/or $G$) and the time variation of the
nuclear and particle physics quantities. Even if the DM does not change
with the cosmic expansion, it is necessary to include it as a part of the
total energy density of the universe.

We assume that the dark matter particles do not vary
with time, i.e. $\dot{m}_X=0$, and only the cosmic evolution of $\LQCD$
accounts for the non-conservation of matter. Trading the cosmic time for
the scale factor through $\dot{\Lambda}_{\rm QCD}=\left(d{\Lambda}_{\rm
QCD}/da\right)\,a\,H$ and integrating the resulting equation, we can
express the final result in terms of the redshift:

\begin{equation}\label{eq:LQCDz}
\LQCD(z)=\LQCD^0\,\left(1+z\right)^{-3\,(\ODMo/\OMBo)\,\nueff}\,.
\end{equation}
For the protons we obtain:
\begin{equation}\label{eq:mN}
m_p(z)=m_p^0\,\left(1+z\right)^{-3\,(\ODMo/\OMBo)\,\nueff}\,.
\end{equation}
Here $\LQCD^0$ and $m_p^0$ are the QCD scale and proton mass at present
($z=0$). $\ODMo$ and $\OMBo$ are the current values of these
cosmological parameters.

The presence of the factor ${\OMBo}/{\ODMo}$ in the
power law makes eq. (\ref{eq:mN}) more realistic than eq.
(\ref{eq:nonconservedmp}). In the case $\nu=0$ the QCD scale and the
proton mass would not vary with the expansion of the universe, but for
non-vanishing $\nu$ it describes the cosmic running of
$\LQCD=\LQCD(z)$ and $m_p=m_p(z)$. For $\nu>0$ ($\nu<0$) the QCD scale and
proton mass decrease (increase) with the redshift. This is consistent, since
for $\nu>0$ ($\nu<0$) the vacuum energy density is
increasing (decreasing) with the redshift -- cf. Eq.\,(\ref{CRG2}) --, and
it is smaller (larger) now than in the past.

We can write down the variation of the QCD scale in terms
of the Hubble rate $H$. With the help of Eq.\,(\ref{nomalHflow}) equation
(\ref{eq:LQCDz}) can be turned into an expression for $\LQCD$ given
explicitly in terms of the primary cosmic variable $H$:
\begin{equation}\label{eq:LQCDH}
\LQCD(H)=\LQCD^0\,\left[\frac{1-\nu}{\OMo}\,\frac{H^2}{H_0^2}-\frac{\OLo-\nu}{\OMo}\right]^{-(\ODMo/\OMBo)\,\nueff/(1-\nu)}\,,
\end{equation}
with $\OMo=\OMBo+\ODMo$. $\nu$ and $\nueff$ are involved in
(\ref{eq:LQCDH}), since they come from different sources. This equation
satisfies the normalization condition $\LQCD(H_0)=\LQCD^0$ due to the
cosmic sum rule for flat space: $\OMo+\OLo=1$.

{Obviously the cosmic time variation of the $\LQCD$ scale is very small in
our framework. This can be more easily assessed if we use
Eqs.\,(\ref{eq:LQCDz}) and (\ref{eq:LQCDH}) to compute the relative time
variation of the QCD scale with respect to the present value. Since $\nu$
is small it it easy to show that}
\begin{equation}\label{eq:deltaLQCDH}
\frac{\LQCD(z)-\LQCD^0}{\LQCD^0}=-\frac{\ODMo}{\OMBo}\,\frac{\nueff}{1-\nu}
\ln\left[\frac{1-\nu}{\OMo}\,\frac{H^2(z)}{H_0^2}-\frac{\OLo-\nu}{\OMo}\right]\,.
\end{equation}
{As a concrete example, let us consider the studies made in
Ref.\,\cite{Reinhold06} on comparing the $H_2$ spectral Lyman and Werner
lines observed in the Q 0347-383 and Q 0405-443 quasar absorption systems.
The comparison with the corresponding spectral lines at present may be
sensitive to a possible evolution of these lines in the last twelve
billion years and involves redshifts in the range $z\simeq 2.6-3.0$. A
positive result could be interpreted as a small variation of the proton to
electron mass ratio (\ref{eq:mupe}) between two widely separated epochs of
the cosmological evolution\,\cite{Reinhold06} . Assuming that $|\nu|={\cal
O}(10^{-3})$, as suggested by Eq.\,(\ref{eq:numodeliicosmology}), it
follows from the previous formulae that the relative variation of $\LQCD$
in this lengthy time interval is only at the few percent level with
respect to its present day value. From Eq.\,(\ref{CRG}) we can then easily
check that the corresponding variation of $\rL(z)$ with respect to the
current value $\rLo$ is also of a few percent. As expected, the two scales
undergo tiny variations over very long periods of time, in fact
cosmological periods, and therefore the large hierarchy between them at
present -- namely $\LQCD={\cal O}(100)$ MeV$={\cal O}(10^8)$ eV and
$\rL^{1/4}={\cal O}(10^{-3})$ eV -- is essentially preserved over the
cosmological evolution. However, even this small crosstalk between these
two widely separated scales could be sufficient for being eventually
detected by the aforementioned high precision experiments aiming at
measuring very tiny variation of the proton to electron mass ratio. This
is suggested by the fact that the expected range of values of $\nu$ is
within the scope of the precision of these experiments. }

Using the above equations and Eq.\,(\ref{alphasLQCD}), we can obtain the
corresponding evolution of the strong coupling constant $\alpha_s$ with
the redshift and the Hubble rate, i.e. $\alpha_s(\mu_R;z)$ and
$\alpha_s(\mu_R;H)$:
\begin{equation}\label{eq:alphasz}
\frac{1}{\alpha_s(\mu_R;z)}=\frac{1}{\alpha_s(\mu_R;0)}+6\,b_0\,
\frac{\ODMo}{\OMBo}\,\nueff\,\ln{(1+z)}\,.
\end{equation}
Here $\alpha_s(\mu_R;0)$ is the value of $\alpha_s(\mu_R;z)$ today
($z=0$). Since $b_0>0$ (cf. sect.\,\ref{sect:AtomicMasses}), we observe
that for $\nu>0$ ($\nu<0$) the strong interaction $\alpha_s(\mu_R;z)$
decreases (increases) with $z$,  i.e. with the cosmic evolution. We also
find\,\footnote{It is interesting to note that a similar running of
$\alpha_s$ with the cosmic expansion was pointed out in a different
context by J.D. Bjorken in \cite{Bjorken2002}.}:
\begin{equation}\label{eq:alphasH}
\frac{1}{\alpha_s(\mu_R;H)}=\frac{1}{\alpha_s(\mu_R;H_0)}+2\,b_0\,
\frac{\ODMo}{\OMBo}\,\frac{\nueff}{1-\nu}\,\ln{\left[\frac{1-\nu}{\OMo}\,\frac{H^2}{H_0^2}-\frac{\OLo-\nu}{\OMo}\right]}\,.
\end{equation}
Here $\alpha_s(\mu_R;H_0)$ is the current value of $\alpha_s(\mu_R;H)$.

Above we have determined the strong coupling as a
function of \textit{two} running scales: one is the ordinary QCD running
scale $\mu_R$, the other is the cosmic scale defined by the Hubble rate
$H$, which has dimension of energy in natural units. The second term on
the \textit{r.h.s.} depends on the product of the two $\beta$-function
coefficients, the one for the ordinary QCD running ($b_0$) and the one for
the cosmic running ($\nu\propto\nueff$).

We find:

\begin{description}

\item i) for $\nu=0$ there is no cosmic running of the strong
    interaction,

\item ii) for $\nu>0$ the strong coupling $\alpha_s(\mu_R;H)$ is
    ``doubly asymptotically
free''. It decreases for large $\mu_R$ and also for large $H$, whereas
for $\nu<0$ the cosmic evolution drives the running of $\alpha_s$
opposite to the normal QCD running,

\item iii) the velocity of the two runnings is very different, because
    $H$ is slowly varying with time and $|\nu|\ll1$ and $|\nu|\ll
    b_0\lesssim 1$. The cosmic running only operates in the cosmic history and is
weighed with a very small $\beta$-function. But it may soon
be measured in the experiments with atomic clocks and through
astrophysical observations.
 \end{description}

The previous equations describe not only the leading cosmic evolution of
the QCD scale and the proton mass with the redshift and the expansion rate
$H$ of the universe, but they can account for the redshift evolution of
the nuclear masses. For the neutron we can write approximately: $m_n\simeq
c_{\rm QCD}\,\LQCD$. For an atomic nucleus of current mass $M_A$ {and
atomic number $A$} we have $M_A=Z\,m_p+(A-Z)\,m_n-B_A$, where $Z$ is the
number of protons and $A-Z$ the number of {neutrons}, and $B_A$ is the
binding energy. {Although $B_A$} may also change with the cosmic
evolution, the shift should be less significant, since at leading order
the binding energy relies on pion exchange among the nucleons. The pion
mass has a softer dependence on $\LQCD$: $m_{\pi}\sim \sqrt{m_q\,\LQCD}$,
due to the chiral symmetry.

In the previous approximations we have neglected the light quark
masses $m_q$. We can assume that the binding energy has a
negligible cosmic shift as compared to the masses of the nucleons. In
the limit where we neglect the proton-neutron mass difference and
assume a common nucleon mass $m_N^0$ at present, the corresponding mass of
the atomic nucleus at redshift $z$ is given at leading order by:
\begin{equation}\label{eq:MAz}
M_A(z)\simeq A\,m_N^0\,\left(1+z\right)^{-3\,(\ODMo/\OMBo)\,\nueff}-B_A\,.
\end{equation}
Although the chemical elements redshift their masses, a disappearance or
overproduction of nuclear mass (depending on the sign of $\nu$) is
compensated by a running of the vacumm energy $\rL$, which is of opposite
in sign, see (\ref{eq:deltaLambda}).

Above we have described a simplified case, in which the nuclear matter
evolves with the cosmic evolution as a result of the evolution
of the fundamental QCD scale. In this scenario the light quark masses are
neglected, and the DM does not participate in the cosmic time evolution.

Alternatively we can assume that the nuclear matter does not vary with
time, i.e. $\dot\Lambda_{\rm QCD}=0$, and only the DM particles account
for the non-conservation of matter. In general we expect a mixed
situation, in which the temporal rates of change for nuclear matter and
for DM particles are different:
\begin{equation}\label{eq:rateNandDM}
\frac{\dot{\Lambda}_{\rm QCD}}{\LQCD}=3\,\nuQCD\,H\,,\ \ \ \ \ \ \ \ \frac{\dot{m}_{X}}{m_X}=3\,\nuX\,H\,.
\end{equation}
We have defined the QCD time variation index, which is characteristic
of the redshift rate of the QCD scale, while $\nuX$ is the corresponding
one for the DM. In this more general case we find:
\begin{equation}\label{eq:LQCDmxDz}
\LQCD(z)=\LQCD^0\,\left(1+z\right)^{-3\,\nuQCD}\,,\ \ \ \ \ \ \ \ m_X(z)=m_X^0\,\left(1+z\right)^{-3\,\,\nuX}\,.
\end{equation}
We introduce the effective baryonic redshift index $\nuB$:
\begin{equation}\label{eq:nuB}
\nuB=\frac{\OMB}{\ODM}\,\nuQCD\,.
\end{equation}
The equations (\ref{eq:LQCDmxDz}) satisfy the relation
(\ref{eq:reltimeMdensityUniv3}), provided the coefficients $\nuB$ and
$\nuX$ are related by
\begin{equation}\label{eq:nunuQCDnuX}
\nueff=\nuB+\nuX\,.
\end{equation}
$\nuQCD$ is the intrisic cosmic rate of variation of the strongly
interacting particles. The effective index $\nu_B$ weighs the redshift
rate of these particles taking into account their relative abundance with
respect to the DM particles. Even if the intrinsic cosmic rate of
variation of $\LQCD$ would be similar to the DM
index  (i.e. if $\nuQCD\gtrsim \nuX)$, the baryonic index (\ref{eq:nuB})
would still be suppressed with respect to $\nuX$, because the total amount
of baryon matter in the universe is much smaller than the total amount of
DM.

In this mixed scenario the mass redshift of the dark matter particles
follows a similar law as in the case of protons (\ref{eq:mN}), except  now
we have $\nueff\rightarrow\nuB$. The proton would have the index $\nuQCD$
characteristic of the free (and bound) stable strongly interacting matter:
\begin{equation}\label{eq:mN2}
m_p(z)=m_p^0\,\left(1+z\right)^{-3\,(\ODM/\OMB)\,\nuB}=m_p^0\,\left(1+z\right)^{-3\nuQCD}\,.
\end{equation}
The DM particles have another independent index $\nuX$. The sum
(\ref{eq:nunuQCDnuX}) must reproduce the original index
$\nueff\propto\nu$, which we associated with the non-conservation of
matter.

Finally we consider the possible quantitative contribution to the
matter density anomaly from the dark matter. The global mass defect (or
surplus) is regulated by the value of the $\nu$ parameter, but the
contribution of each part (baryonic matter and DM) depends on the values
of the individual components $\nuB$ and $\nuX$. We can obtain a numerical
estimate of these parameters by setting the expression
(\ref{eq:reltimeMdensityUniv}) equal to (\ref{eq:deltadotrho2}). The
latter refers to the time variation of the matter density $\rM$ without
tracking the particular way in which the cosmic evolution can generate an
anomaly in the matter conservation. The former does assume that
this anomaly is entirely due to a cosmic shift in the masses of the stable
particles. Taking the absolute values, we obtain:
\begin{equation}\label{eq:equatetwodensit}
3|\nueff|\,H\simeq \left|\frac{4}{23}\,\frac{\dot{\Lambda}_{\rm QCD}}{\LQCD}+
\frac{\dot{m}_X}{m_X}\right|< \frac{4}{23}\times 10^{-14}\,{\rm yr}^{-1}+\left|\frac{\dot{m}_X}{m_X}\right|\,.
\end{equation}
Here we have used the experimental bound (\ref{eq:timeLambda}) on the time
variation of $\LQCD$.

Several cases can be considered, depending on the relation between the
intrinsic cosmic rates of variation of the strongly interacting particles
and DM particles, $\nuQCD$ and $\nuX$. Since these indices can have either
sign, we shall compare their absolute values:

\begin{itemize}

\item 1) $|\nuX|\ll|\nuB|$:

This condition implies $|\nuX|\ll|\nuQCD|$. By demanding the stronger
condition $|\nuX|\ll|\nuB|$, we insure that the intrinsic QCD cosmic
rate $|\nuQCD|$ is much larger than the corresponding DM rate
$|\nuX|$. We can neglect the $\dot{m}_X/m_X$ term on the
\textit{r.h.s.} of (\ref{eq:equatetwodensit}), and we recover the
equations (\ref{eq:LQCDz})-(\ref{eq:alphasH}) with $\nueff\simeq\nuB$.
Using $H_0\simeq 7\times 10^{-11}\,{\rm yr}^{-1}$\,, we find:
\begin{equation}\label{eq:scenario1a}
|\nuX|\simeq 0\,,\ \ \ \ \ |\nueff|\simeq|\nuB|< 10^{-5}\,,\ \ \ \ |\nuQCD|<5\times 10^{-5}\,.
\end{equation}
The bound on $\nuB\simeq\nueff$ that we have obtained above can be
compared with (\ref{nuboundfromLab}). The former (which is more
stringent) is more realistic than the latter because here we have
taken into account explicitly the suppression factor $\OMB/\ODM$ of
baryonic matter versus dark matter -- and also the (small) difference
between $\nu$ and $\nueff$.
\item 2) $|\nuX|\simeq|\nuB|$:

Here we still have $|\nuX|$ smaller than $|\nuQCD|$, but the
requirement is weaker. It follows: $|\nueff|\simeq2|\nuX|\simeq
2|\nuB|=2(\OMB/\ODM)\,|\nuQCD|$, and we find
\begin{equation}\label{eq:scenario2}
|\nueff|<2\times 10^{-5}\,,\ \ \ \ \ |\nuX|\simeq |\nuB|<10^{-5}\,,\ \ \ \ \ \ \ \ |\nuQCD|<5\times 10^{-5}\,.
\end{equation}
\item 3) $|\nuX|\simeq|\nuQCD|$:

The two intrinsic cosmic rates for strongly interacting and DM
    particles are similar, i.e. $\dot{\Lambda}_{\rm QCD}/\LQCD$ and
    $\dot{m}_X/m_X$ do not differ significantly.  In this case
 Eq.\,(\ref{eq:equatetwodensit}) leads to
\begin{equation}\label{eq:scenario2b}
3|\nueff|\,H<\left(\frac{4}{23}+ 1\right)\times 10^{-14}\,{\rm yr}^{-1}\,.
\end{equation}
There are two sign possibilities ($\nuQCD=\pm\nuX$), and we take the
absolute value:
\begin{equation}\label{eq:nuQCDeqnuX}
|\nueff|\lesssim\left(\frac{\OMB}{\ODM}+ 1\right)\,|\nuQCD|\simeq |\nuQCD|\,.
\end{equation}
We find:
\begin{equation}\label{eq:scenario2c}
|\nueff|\lesssim|\nuQCD|\simeq |\nuX|<5\times 10^{-5}\,.
\end{equation}
\item 4) $|\nuQCD|\ll|\nuX|$:

Here the nuclear part is frozen. The non-conservation of matter is
entirely due to the time variation of the DM particles.
Eq.\,(\ref{eq:equatetwodensit}) gives:
\begin{equation}\label{eq:scenario3}
3\nu\,H\simeq \frac{\dot{m}_X}{m_X}\,\left(1-\frac{\OMB}{\ODM}\right)\,.
\end{equation}
\begin{table*}[t]
\tabcolsep 3pt \vspace {0.2cm}
\begin{center}
\begin{tabular}{|c||l|c|c|}
  \hline
   &\phantom{XXXXX} $|\nu|^{\rm cosm}$ & $|\nu|^{\rm lab}=|\nuB|$ & $|\nuX|^{\rm cosm}$ \\ \hline\hline
  {\rm Model II} & $10^{-3}$ (SNIa+BAO+CMB)& $10^{-5}$ (Atomic clocks+Astrophys.) & $10^{-3}$ \\ \hline
   {\rm Model III} & $10^{-3}$ (BBN)& $10^{-5}$ (Atomic clocks+Astrophys.) & $10^{-3}$ \\ \hline
   {\rm Model IV} & $10^{-3}$ (SNIa+BAO+CMB)+BBN& 0 & 0 \\
  \hline
\end{tabular}
\caption[]{Upper bounds on the running index $|\nu|$ for the various
models defined in sect.\,\ref{sect:timeCC}. Only for Models II and III
a non-vanishing value of $|\nu|$ is related to non-conservation of
matter and a corresponding time evolution of $\rL$ and $G$,
respectively. For these models, a part of $\nu$ (viz. $\nuB$) is
accessible to lab experiments, whereas the DM contribution ($\nuX$)
can only be bound indirectly from cosmological observations (same
cosmological bound as for the overall $\nu$). For Model IV matter is
conserved, and a non-vanishing value of $|\nu|$ (only accessible from
pure cosmological observations) is associated to a simultaneous time
evolution of $\rL$ and $G$ -- with no microphysical implications.}
\end{center}
\end{table*}
We have written this expression directly in terms of the original
$\nu$ parameter. In this case we cannot get information from any
laboratory experiment on $\dot{m}_X/m_X$, but we do have independent
experimental information on the original $\nu$ value (irrespective of
the particular contributions form the nuclear and DM components). It
comes from the cosmological data on type Ia supernovae, BAO, CMB and
structure formation. The analysis of this data\,\cite{BPS09,GSBP11}
leads to the bound (\ref{eq:numodeliicosmology}), which applies to all
models, in which matter follows the generic non-conservation law
(\ref{mRG2}) and the running vacuum law (\ref{RGlaw2}) --- or the same
matter non-conservation law and the running gravitational coupling law
(\ref{GzfixedLsmallnu}), as shown in Eq.\,(\ref{eq:nuboundGvariable}).
Since it depends on the cosmological effects from all forms of matter,
it applies to the DM particles in particular. We find:
\begin{equation}\label{eq:scenario3b}
|\nuX|^{\rm cosm}\lesssim 10^{-3}\,.
\end{equation}
This bound is significantly weaker than any of the bounds found for
the previous scenarios in which the nuclear matter participated of the
cosmic time variation. It cannot be excluded that the matter
non-conservation and corresponding running of the vacuum energy in the
universe is mainly caused by the general redshift of the DM particles.
In this case only cosmological experiments could be used to check this
possibility. If the nuclear matter also participates in a significant
way, it could be analyzed with the help of experiments in the
laboratory. For a summary of the bounds, see Table 1.

\end{itemize}

If in the future we could obtain a tight cosmological bound on the
effective $\nueff$-parameter (\ref{eq:nunuQCDnuX}), using the
astrophysical data, and an accurate laboratory (and/or astrophysical)
bound on the baryonic matter part $\nuB$, we could compare them and derive
the value of the DM component $\nuX$. If $\nueff$ and $\nuB$ would be
about equal, we should conclude that the DM particles do not appreciably
shift their masses with the cosmic evolution, or that they do not exist.
If, in contrast, the fractional difference $|\,(\nueff-\nuB)/\nueff\,|$
would be significant, the DM particles should exist to compensate for it.

\mysection{Conclusions}\label{sect:conclusions}

In this paper we have described theoretical models based on the
assumption that the basic constants of nature are slowly varying
functions of the cosmic expansion, as suggested by numerous
experiments. We have connected the variation of the nuclear and
particle masses, fundamental scales and particle physics couplings
(e.g. the fine structure and the strong coupling ``constant'') to
the possible cosmic evolution of the two parameters $\rL$ and $G$ of
Einstein's gravity theory, i.e. the vacuum energy density (or
cosmological ``constant'') and the gravity constant. The
non-conservation of matter, associated to a time variation of the
parameters in nuclear and particle physics, must be compensated by
the corresponding evolution of the vacuum energy density and/or
gravitational coupling $G$. This resulting picture of the cosmic
evolution is compatible with the cosmological principle, but $\rL$
and/or $G$ evolve with the cosmic time in combination with the
fundamental ``constants''\,\cite{Constants}.

We have represented the possible time evolution of the physical quantities
in terms of the effective (dimensionless) parameter $\nueff$, proportional
to the original $\nu$. If the experiments would detect a mass density
anomaly in the microphysics world, e.g. through a (red)shift in the value
of the proton to electron mass ratio, it would lead to a non-vanishing
value of $\nuB$ (which is the baryonic part of $\nueff$). This anomaly
would be correlated with the corresponding shift of the dimensionless
quantities $\CC/m_p^2$ and $G\,m_p^2$  (for the class of Models II and III
respectively). A shift in the value of these dimensionless quantities
would determine $\nu\propto\nueff$, and the corresponding value of
$\nueff$ could be confronted with a possible mass anomaly $\nuB$ of the
nuclear matter. From the difference with $\nueff$ we could infer an
indirect effect from dark matter, which is controlled by the dimensionless
index $\nuX$.

We have described the cosmic evolution of the various quantities through
the Hubble rate $H$ as the basic scale, which can parameterize the running
of the masses and couplings as well as the vacuum energy and/or Newton's
constant $G$. The running of $\rL$ and $G$ is related to the quantum
effects of the particles on the effective action of  QFT in curved
space-time. The vacuum energy density is written as a function of $H$:
$\rL=\rL(H)$.

Matter is non-conserved. We have attributed the non-conservation to a
cosmic redshift (hence a cosmic time variation) of the masses of the
nucleons, due to the corresponding change of the QCD parameter $\LQCD$.
All atoms would be affected as well.  One may expect that the redshift
should affect the masses of all the fundamental particles (quarks and
leptons), including the dark matter particles. We have explicitly proposed
a connection of the cosmic time evolution of the $\LQCD$ scale and of the
elementary particle masses to the corresponding running of $\rL$ and/or
$G$.

The present bounds, obtained for the time variation of the fundamental
constants of nuclear matter, point to a rate of change of the nucleon mass
and the $\LQCD$ parameter, which is compatible with the corresponding
bounds on the cosmic evolution of  $\rL$ and $G$. The relevant
dimensionless parameter, which controls the running of these quantities,
must be of order $|\nu|\lesssim10^{-3}$ or less. The current time
variation of the vacuum energy can be of order
$\left|\dot\rho_{\CC}/\rL\right|\sim 10^{-3}\,H_0\sim 10^{-14}\, {\rm
yr}^{-1}$. This can be compared with the current measured rate of change
of the $\LQCD$ scale in astrophysical and in atomic clock experiments,
which provide bounds of the same order of magnitude. However the
laboratory  bounds affect only the nuclear matter contribution to $\nu$.
The remaining contribution, as indicated before, should come from dark
matter particles. This approach could eventually provide an indirect
evidence, that dark matter particles exist.

{Let us clarify that in our framework we cannot provide at this stage an
explanation for the value of the cosmological constant nor of the QCD
scale. The mass scale associated to the vacuum energy or DE is
$m_{\CC}\equiv\left(\rLo\right)^{1/4}\sim 10^{-3}$ eV, which is roughly
eleven orders of magnitude smaller than the value of the QCD scale,
$\LQCD=O(100)$ MeV. To explain the former from first principles would be
so much as to provide a solution of the old cosmological constant problem,
whereas to explain the latter would be tantamount to explain quark
confinement. We do not provide here a clue for the solution of any of
these  problems, but we suggest that there may be a crosstalk between the
scale of the measured vacuum energy in cosmology and the scale of the
strong interactions (and in general with the particle masses). Despite we
do not understand at this point the absolute value of these scales, the
possible interaction between them could smoothly shift their values with
the cosmic time; and this cosmological evolution could be measured both at
the astrophysical level and in the laboratory within the next generation
of atomic clocks\,\cite{Haensch12}.}

The models of the cosmic evolution, discussed in this paper, offer an
interesting perspective to unify the microphysical and the {macrophysical}
laws of nature. The dark energy is the dynamical vacuum energy in
interaction with matter. If the dark matter would participate in the
cosmic redshift, affecting the baryonic matter, there would be an intimate
connection between the evolution of the dark matter and of the dark
energy.

The ideas presented here can be tested by different kind of experiments.
They could help to understand the structure and cosmic behavior of
ordinary matter as well as to uncover the mysteries of dark matter and
dark energy. The small cosmic variation of the physical
``constants'' may signal a connection between
the large scale structure of the universe and
the quantum phenomena in the microcosmos.

\vspace{0.5cm}

\noindent{\bf Acknowledgments} \vspace{0.2cm}

\noindent HF would like to thank Prof. Phua from the Institute for
Advanced Study at the Nanyang Technological University in Singapore for
support. JS has been supported in part by MEC and FEDER under project
FPA2010-20807, by the Spanish Consolider-Ingenio 2010 program CPAN
CSD2007-00042 and by DIUE/CUR Generalitat de Catalunya under project
2009SGR502. He is grateful to D. Pavon for pointing out an interesting
reference.

\newcommand{\JHEP}[3]{ {JHEP} {#1} (#2)  {#3}}
\newcommand{\NPB}[3]{{ Nucl. Phys. } {\bf B#1} (#2)  {#3}}
\newcommand{\NPPS}[3]{{ Nucl. Phys. Proc. Supp. } {\bf #1} (#2)  {#3}}
\newcommand{\PRD}[3]{{ Phys. Rev. } {\bf D#1} (#2)   {#3}}
\newcommand{\PLB}[3]{{ Phys. Lett. } {\bf B#1} (#2)  {#3}}
\newcommand{\EPJ}[3]{{ Eur. Phys. J } {\bf C#1} (#2)  {#3}}
\newcommand{\PR}[3]{{ Phys. Rep. } {\bf #1} (#2)  {#3}}
\newcommand{\RMP}[3]{{ Rev. Mod. Phys. } {\bf #1} (#2)  {#3}}
\newcommand{\IJMP}[3]{{ Int. J. of Mod. Phys. } {\bf #1} (#2)  {#3}}
\newcommand{\PRL}[3]{{ Phys. Rev. Lett. } {\bf #1} (#2) {#3}}
\newcommand{\ZFP}[3]{{ Zeitsch. f. Physik } {\bf C#1} (#2)  {#3}}
\newcommand{\MPLA}[3]{{ Mod. Phys. Lett. } {\bf A#1} (#2) {#3}}
\newcommand{\CQG}[3]{{ Class. Quant. Grav. } {\bf #1} (#2) {#3}}
\newcommand{\JCAP}[3]{{ JCAP} {\bf#1} (#2)  {#3}}
\newcommand{\APJ}[3]{{ Astrophys. J. } {\bf #1} (#2)  {#3}}
\newcommand{\AMJ}[3]{{ Astronom. J. } {\bf #1} (#2)  {#3}}
\newcommand{\APP}[3]{{ Astropart. Phys. } {\bf #1} (#2)  {#3}}
\newcommand{\AAP}[3]{{ Astron. Astrophys. } {\bf #1} (#2)  {#3}}
\newcommand{\MNRAS}[3]{{ Mon. Not. Roy. Astron. Soc.} {\bf #1} (#2)  {#3}}
\newcommand{\JPA}[3]{{ J. Phys. A: Math. Theor.} {\bf #1} (#2)  {#3}}
\newcommand{\ProgS}[3]{{ Prog. Theor. Phys. Supp.} {\bf #1} (#2)  {#3}}
\newcommand{\APJS}[3]{{ Astrophys. J. Supl.} {\bf #1} (#2)  {#3}}

\newcommand{\Prog}[3]{{ Prog. Theor. Phys.} {\bf #1}  (#2) {#3}}
\newcommand{\IJMPA}[3]{{ Int. J. of Mod. Phys. A} {\bf #1}  {(#2)} {#3}}
\newcommand{\IJMPD}[3]{{ Int. J. of Mod. Phys. D} {\bf #1}  {(#2)} {#3}}
\newcommand{\GRG}[3]{{ Gen. Rel. Grav.} {\bf #1}  {(#2)} {#3}}




\begin{thebibliography}{50}


\bibitem{HiggsFinding} ATLAS Collaboration, arXiv:1202.1408;
    arXiv:1202.1414; arXiv:1202.1415; arXiv:1202.1636; CMS Collaboration, arXiv:1202.1416;
    arXiv:1202.1487; arXiv:1202.1488; arXiv:1202.1489; arXiv:1202.1997.

\bibitem{WMAP} E.~Komatsu {\it et al.} [WMAP Collaboration],
\APJS {180}{2009}{330} [arXiv:0803.0547];
\APJS {192}{2011}{18} [arXiv:1001.4538].

\bibitem{SNIa} R.~Knop { et al.},
\APJ {598}{2003}{102} [arXiv:astro-ph/0309368];
%
A.~Riess \textit{et al.}
\APJ {607}{2004}{665} [arXiv:astro-ph/0402512].


\bibitem{WeinbergRMP} S. Weinberg, \RMP {\bf 61} {1989}  {1}.

\bibitem{PeeblesRatra03} P.J.E.~Peebles and B.~Ratra, \RMP
    {75}{2003}{559}.

\bibitem{CCRev} V. Sahni, A. Starobinsky, \IJMP {A9} {2000} {373}; S.M.
    Carroll, \textsl{Living Rev. Rel.} {\bf 4} (2001) 1; T. Padmanabhan,
    \PR {380} {2003} {235}; E.J. Copeland, M. Sami, S. Tsujikawa, \IJMP
    {D15} {2006} {1753}; M. Li , X-D. Li, S. Wang, Y. Wang, Commun. Theor. Phys. {\bf 56} (2011) 525
    [arXiv:1103.5870].

\bibitem{Zeldovich67} {Y. B. Zeldovich}, \textit{Cosmological constant and
    elementary particles}, Sov. Phys. JETP Lett {\bf 6} (1967) {316};
    Soviet Physics Uspekhi {\bf 11} (1968) 381.

\bibitem{RelaxTH} F.~Bauer, J.~Sol\`a, H.~\v{S}tefan\v{c}i\'{c}, %
JCAP {\bf 12 } (2010) 029 [arXiv:1006.3944].

\bibitem{FritzschBarcelona11} H. Fritzsch, \textit{Fundamental constants
    and their time variation}, invited talk at Univ. of Barcelona,
    November 2011.


\bibitem{Fritzsch11} H. Fritzsch, \textit{Fundamental constants and their
    time variation}, Prog. Part. Nucl. Phys. {\bf 66} (2011) 193;
    \textit{The fundamental constants in physics and their possible time
    variation}, Nucl.Phys. Proc. Suppl. {\bf 203-204} (2010) 3.

\bibitem{Fritzsch06}
X. Calmet, H. Fritzsch, Europhys. Lett. {\bf 76} (2006) 1064
[arXiv:astro-ph/0605232];
\PLB {540}{2002}{173} [arXiv:hep-ph/0204258];
\EPJ {24}{2002}{639} [arXiv:hep-ph/0112110].

\bibitem{Langacker02}
P. Langacker, G. Segre, M. J. Strassler, \PLB {528}{2002}{121}
[arXiv:hep-ph/0112233];
V.V. Flambaum, D.B. Leinweber, A.W. Thomas, and R.D. Young, \PRD
{69}{2004}{115006} [arXiv:hep-ph/0402098].


\bibitem{Shlyakhter1976}  A.I. Shlyakhter, Nature 264 (1976) 340.

\bibitem{DamourDyson96}
T. Damour, F. Dyson, \NPB {480}{1996}{37} [arXiv:hep-ph/9606486].

\bibitem{Fujii2000} Y. Fujii, A. Iwamoto, T. Fukahori, T. Ohnuki, M.
    Nakagawa, H. Hidaka, Y. Oura, and P. M\"oller, \NPB {573}{2000}{377}
    [arXiv:hep-ph/9809549]; [arXiv:hep-ph/0205206].


\bibitem{Haensch12} T.W. H\"ansch, private communication.

\bibitem{Haensch04}
M. Fischer, N. Kolachevsky, M. Zimmermann, R. Holzwarth, T. Udem, T.W.
H\"ansch  et al, \PRL {92}{2004}{230802}.


\bibitem{AtomicClocks}
S. Blatt, et al., \PRL {100}{2008}{140801} [arXiv:0801.1874];
S. Bize, et al.. \PRL {90}{2003}{150802}.


\bibitem{Reinhold06}
E. Reinhold, R. Buning, U. Hollenstein, A. Ivanchik, P. Petitjean, W.
Ubachs, \PRL {96}{2006}{151101};
W. Ubachs, E. Reinhold, \PRL {92}{2004}{101302};
A. Ivanchik, P. Petitjean, D. Varshalovich, B. Aracil, R. Srianand, H.
Chand, C. Ledoux, P. Boisse, Astron. Astrophys. {\bf 440} (2005) 45
[arXiv:astro-ph/0507174].

\bibitem{JSP11a} J.~Sol\`a, \textit{Cosmologies with a time dependent
    vacuum}, {J. Phys. Conf. Ser.} {\bf 283} (2011) 012033
    [arXiv:1102.1815];
Fortsch. Phys. {\bf 59} (2011) 1108, and references therein.

\bibitem{Dirac37} P.A.M. Dirac, Nature 139 (1937) 323.

\bibitem{MilneJordan} E. A. Milne, Relativity, Gravitation and World
    Structure (Clarendon press, Oxford, 1935); Proc. Roy. Soc. {\bf A3}
    (1937) 242; P. Jordan, Naturwiss. {\bf 25} (1937) 513; Z. Physik {\bf
    113} (1939) 660.

\bibitem{JBD} P. Jordan, \textit{Nature} {\bf 164} (1949) 637; C. Brans,
    R.H. Dicke, \PRD {124}{1961}{925}.

\bibitem{CCvariable} See e.g. the reviews:
J. M. Overduin and F. I. Cooperstock, \PRD {58} {1998} {043506}; and
R.G. Vishwakarma,  \CQG {18} {2001} {1159}, and references therein.


\bibitem{OldScalar} A.D.~Dolgov, in: \textit{The very Early Universe}, Ed.
    G.~Gibbons, S.W.~Hawking, S.T.~Tiklos (Cambridge U., 1982); {L.F.
    Abbott}, \PLB {150}{1985}{427}; L.H.~Ford, \PRD {35}{1987}{2339};
    R.D.~Peccei, J.~Sol\`{a} and C.~Wetterich, \PLB {195}{1987}{183}; S.
    M. Barr, \PRD {36}{1987}{1691}; J.~Sol\`{a}, \PLB {228}{1989}{317};
    \IJMP {A5}{1990}{4225}.

\bibitem{Uzan11} J-P. Uzan,
Living Rev. Rel. {\bf 14} (2011) 2 [arXiv:1009.5514];
\RMP {75}{2003}{403} [arXiv: hep-ph/0205340].

\bibitem{Chiba11}
T. Chiba, Prog. Theor. Phys. 126 (2011) 993 [arXiv:1111.0092].


\bibitem{Murphy03}
M.T. Murphy, J.K. Webb, V.V. Flambaum, \MNRAS {345}{2003}{609}
[arXiv:astro-ph/0306483];
M.T. Murphy, J.K. Webb and V.V. Flambaum, Lec. Not. Phys. 648 (2004) 131
[arXiv:astro-ph/0310318];
J. K. Webb et al., \PRL {87}{2001}{091301} [arXiv:astro-ph/0012539].

\bibitem{No-alphateffect}
H. Chand, R. Srianand, P. Petitjean, and B. Aracil, Astron. and Astroph.
{\bf 417}  (2004) 853 [arXiv:astro-ph/0401094];
R. Quast, D. Reimers, and S.A. Levshakov, Astron. and Astroph. {\bf 415}
(2004) 27 [arXiv:astro-ph/0311280].

\bibitem{Fossil07} J. Sol\`a, {J. of Phys.} {\bf A41} {(2008)} {164066}
    [arXiv:0710.4151].

\bibitem{SS09} I. L.  Shapiro, and J. Sol\`a,
Phys. Lett. {\bf B682} (2009) 105 [arXiv:0910.4925]; \\
c.f. also the extended  version arXiv:0808.0315, and references therein;
JHEP {\bf 02} (2002) 006 [hep-th/0012227];  I. L. Shapiro, J. Sol\`a,
H.~\v{S}tefan\v{c}i\'{c}, \JCAP
    {01} {2005} {012} [arXiv:hep-ph/0410095]; J. Sol\`a, H. \v{S}tefan\v{c}i\'{c}, \PLB {624}{2005}{147} [arXiv:
astro-ph/0505133]; \MPLA {21} {2006} {479} [arXiv:astro-ph/0507110].


\bibitem{Parker09} {L.E. Parker} and D.J. Toms, \textit{Quantum
    Field Theory in Curved Spacetime: quantized fields and gravity}
    (Cambridge U. Press, 2009).


\bibitem{BPS09} S. Basilakos, M. Plionis and J. Sol\`a, Phys. Rev. {\bf
    D80} (2009) {3511} [arXiv:0907.4555].


\bibitem{Cmunoz} C. Mu\~noz, private communication (JS thanks him for this
    information).

\bibitem{stringscale}
L. E. Ibanez, C. Mu\~noz, S. Rigolin, \NPB {553}{1999}{43}
[arXiv:hep-ph/9812397].


\bibitem{GSBP11} J. Grande, J. Sol\`a, S. Basilakos, and M. Plionis, \JCAP
    {08} {2011} {007} [arXiv:1103.4632].


\bibitem{BPS10} S. Basilakos, M. Plionis and J. Sol\`a,  Phys.
    Rev. D82 (2010) 083512 [arXiv:1005.5592].

\bibitem{BSP12a} S. Basilakos, D. Polarski and J. Sol\`a, Phys. Rev. {\bf
    D86} (2012) 043010 [arXiv:1204.4806].


\bibitem{RelaxObsvandLXCDM} S. Basilakos, F. Bauer, and J.~Sol\`a,
\JCAP {01}{2012}{050} [arXiv:1109.4739]; F. Bauer, J.~Sol\`a,
H.~\v{S}tefan\v{c}i\'{c}, \MPLA {26} {2011} {2559} [arXiv:1105.1030]; J.
Grande, A. Pelinson, J. Sol\`{a}, \PRD{79}{2009}{043006}
[arXiv:0809.3462].




\bibitem{Mohr2008}
 P.J. Mohr, and B.N. Taylor, \RMP {80}{2008}{633} [arXiv:0801.0028].

 \bibitem{King2008}
 J. King, J.K. Webb, M.T. Murphy, R.F. Carswell, \PRL {101}{2008}{251304} [arXiv:astro-ph/0807.4366].


 \bibitem{Guberina2006}
B. Guberina, R. Horvat, H. Nikolic, \PLB {636}{2006}{80}
[astro-ph/0601598].

\bibitem{oldCCstuff1} I.L. Shapiro, J. Sol\`a, C. Espa\~na-Bonet, P.
    Ruiz-Lapuente,  \PLB {574} {2003} {149} [arXiv:astro-ph/0303306];
    \JCAP {02} {2004} {006} [arXiv:hep-ph/0311171].

\bibitem{scalingmatter2} P. Wang and X. Meng, Class. Quant. Grav {\bf 22}
(2005) 283 [arXiv:astro-ph/0408495]; \\
J. S. Alcaniz and J. A. S. Lima, Phys. Rev. {\bf D72} (2005) 063516
[arXiv:astro-ph/0507372].

\bibitem{GSFS10} J. Grande, J. Sol\`a, J.C. Fabris, I.L.Shapiro, {Class.
    Quant. Grav.} {\bf 27} (2010) 105004 [arXiv:1001.0259].

\bibitem{Iocco:2008va}
F.~Iocco, G.~Mangano, G.~Miele, O.~Pisanti and
    P.~D.~Serpico,
  Phys.\ Rept.\  {\bf 472} (2009) 1
  [arXiv:0809.0631].

 \bibitem{Bjorken2002} 
J. D. Bjorken, \PRD {67}{2003}{043508} [hep-th/0210202].

\bibitem{Constants} 
J. D. Barrow, J. Magueijo, \PLB {443}{1998}{104} [arXiv:astro-ph/9811072];
M.J. Duff, L.B. Okun, G. Veneziano, \JHEP {03}{2002}{023}
[physics/0110060];
H. B. Sandvik, J. D. Barrow, J. Magueijo, \PRL {88}{2002}{031302}
[arXiv:astro-ph/0107512];
\PRD {65}{2002}{063504} [arXiv:astro-ph/0109414];
P. C. Davies, T. M. Davis and C. H. Lineweaver, Nature {\bf 418}  (2002)
602;
M.J. Duff [hep-th/0208093];
J.W. Moffat [hep-th/0208109];
J. D. Bekenstein, \PRD {66}{2002}{123514} [arXiv:gr-qc/0208081];
J. Rich,  Am. J. Phys. {\bf 71} (2003) 1043 [physics/0209016];
P. Langacker, \IJMP {A19S1} {2004} {157} [hep-ph/0304093];
F. Wilczek [arXiv:0708.4361];
A. Moss, A. Narimani, D. Scott. \IJMP {D19} {2010} {2289}
[arXiv:1004.2066].










\end{thebibliography}
\end{document}